\documentclass[11pt]{article}
\usepackage{jcappub}
\usepackage{aas_macros}
\usepackage{mathrsfs}
\usepackage{xspace}

\newlength{\wdo}
\newcommand{\stroke}[1]{{$#1$}%
\settowidth{\wdo}{${#1}$} {\kern-\wdo}%
\partialvartstrokedint}

\makeatletter
\newcommand{\fancysep}{%
  \@afterindentfalse
  {\begin{center}
    \resizebox{0.8\linewidth}{0.4ex}{{%
        \fontsize{20}{24}\usefont{U}{webo}{xl}{n}{4}}}%
  \end{center}}\@afterheading}
\makeatother

{\vskip 1.0em
\framebox[\columnwidth][r]{%
\begin{minipage}[c]{\columnwidth}%
\vspace{-1.0em}%
#1%
\end{minipage}}}
{\vskip 1.0em}

\def\XXint#1#2#3{{\setbox0=\hbox{$#1{#2#3}{\int}$}
     \vcenter{\hbox{$#2#3$}}\kern-.5\wd0}}

\newcommand{\beq}{\begin{equation}}
\newcommand{\eeq}{\end{equation}}
\newcommand{\beqa}{\begin{eqnarray}}
\newcommand{\eeqa}{\end{eqnarray}}

\newcommand{\jHeiv}{$^4$He}

\newcommand{\flrw}{Friedmann-Lema\^{i}tre-Robertson-Walker}

\newcommand{\neff}{\ensuremath{N_{\mbox{\scriptsize eff}}}\xspace}
\newcommand{\lcdm}{\ensuremath{\Lambda}CDM}
\newcommand{\heii}{\mbox{\tiny He {\sc ii}}}
\newcommand{\heiii}{\mbox{\tiny He {\sc iii}}}

\newcommand{\barnum}{\ensuremath{\omega_b}\xspace}
\newcommand{\summnu}{\ensuremath{\sum m_\nu}\xspace}
\newcommand{\neffth}{\ensuremath{N_{\mbox{\scriptsize eff}}^{\mbox{\scriptsize (th)}}}\xspace}
\newcommand{\neffobs}{\ensuremath{\tilde{N}_{\mbox{\scriptsize eff}}}\xspace}
\newcommand{\lnu}{\ensuremath{L_\nu}\xspace}
\newcommand{\numr}{$\nu$MR\xspace}

\graphicspath{{./new_plots_v3/}{./}}

\font\FermiSmallfont=cmssq8 scaled 1200

\def\LANLppthead#1{
\null 
\begin{center}\vskip -1.0truein{\hbox to 7.0truein {
\hfill
\vbox to 1in {\vfill \FermiSmallfont
              \hbox{#1}
              \vfill}
}}\vskip-0.0truein\end{center}}

\title{Probing neutrino physics with a self-consistent
treatment of the weak decoupling, nucleosynthesis, and
photon decoupling epochs}

\author[a,1]{E.\ Grohs,\note{Corresponding author.}}
\author[a]{George M.\ Fuller,}
\author[a,b]{Chad T.\ Kishimoto,}
\author[c]{and Mark W.\ Paris}

\affiliation[a]{Department of Physics, University of California, San Diego, La
Jolla, California 92093-0319, USA}
\affiliation[b]{Department of Physics, University of San Diego, San Diego,
California 92110, USA}
\affiliation[c]{T-2 Nuclear \& Particle Physics, Astrophysics \& Cosmology,
Theoretical Division, Los Alamos National Laboratory, Los Alamos, New
Mexico 87545, USA}

\emailAdd{egrohs@physics.ucsd.edu}
\emailAdd{gfuller@ucsd.edu}
\emailAdd{ckishimo@physics.ucsd.edu}
\emailAdd{mparis@lanl.gov}

\abstract{We show that a self-consistent and coupled treatment of the
   weak decoupling, big bang nucleosynthesis, and photon decoupling
   epochs can be used to provide new insights and
   constraints on neutrino sector physics from high-precision
   measurements of light element abundances and cosmic microwave
   background observables.
   Implications of beyond-standard-model physics in cosmology,
   especially within the neutrino sector, are assessed by comparing
   predictions against five observables: the baryon energy density,
   helium abundance, deuterium abundance, effective number of
   neutrinos, and sum of the light neutrino mass eigenstates.
   We give examples for constraints on dark radiation, neutrino rest
mass, lepton numbers, and scenarios for light and heavy sterile
neutrinos.  }


\begin{document}

\LANLppthead{LA-UR-15-20748}

\maketitle
\flushbottom

\section{Introduction}
\label{sec:intro}
In this paper we demonstrate how a self-consistent treatment of the
physics of the early universe (from temperature $T\sim10$ MeV down to
$T\sim0.1$ eV) enables cosmic microwave background (CMB)
observations, in concert with light-element
abundances, to be used as new probes of beyond-standard-model (BSM)
physics in the neutrino sector.  In a sense, these probes are
tantamount to probes of the C$\nu$B (relic Cosmic Neutrino
Background).  Recent observations
\cite{WMAP:2013ny,ACT:2013cp,SPT:2011hl,PlanckI:2013ov} of the 
CMB and observationally-inferred
primordial abundances of deuterium and helium \cite{Izotov:2010ns,
Pettini:2012yd,Aver:2013ue,Cooke:2014do} formed in big bang
nucleosynthesis (BBN) already place tight constraints on both the
cosmological standard model (CSM) and BSM physics.  However, future
observations will usher in a higher level of precision with even
better prospects for BSM probes,
see for example Ref.\ \cite{Shimon:2010cb}.

We anticipate that observations will bring about an overdetermined
situation where BSM physics may manifest itself if it is present.
The existing or anticipated observations and measurements of greatest
utility for the present purposes are: (1) high-precision measurements
of the baryon-to-photon ratio, or the equivalent baryon density
($\barnum\equiv\Omega_bh^2$); (2) high-precision measurements of the 
``effective number of relativistic degrees of freedom'' (\neff); (3)
high-precision measurements of the primordial deuterium abundance
(D/H) from quasar absorption lines; (4) measurements of the primordial
helium abundance ($Y_P$) directly from CMB polarization data; and (5)
measurements of the sum of the light neutrino masses (\summnu), i.e.\
the collisionless damping scale associated with neutrinos.

The physics that determines the relic neutrino energy spectra in weak
decoupling and that of primordial nucleosynthesis affects observables
of the CMB. These distinct, disparate epochs, however, depend not only
on the values of parameters describing the cosmology (such as
$\omega_b$, $\neff$, relevant cosmic constituents, etc.) but also on
the parameters derived from these base, input parameters (such as
primordial abundances, recombination history, etc.). This is the
basis for what we will term ``self-consistency.'' The requirement for
self-consistency between the BBN and CMB epochs, for example, in
current data analyses depends on the parametrization of the energy
density in terms of \neff\ and upon the baryon-to-photon ratio. These
two parameters, as measured at photon decoupling, are the sole
determinants of the primordial helium abundance at the epoch of alpha
particle formation ($T\sim0.1$ MeV) in ``standard'' ({\em i.e.}\ zero
lepton numbers, no BSM physics) BBN calculations.  Though this
procedure is adequate for the CSM and standard model physics, here we
argue that it can be insufficient to probe varieties of BSM physics when
confronting model cosmologies with next-generation, high-precision CMB
and light element abundance data.  In the case of the helium and \neff\
self-consistency mentioned above, the relationship between the helium
yield in BBN and \neff\ is a non-trivial function of the interplay of
expansion rate and neutron-to-proton ratio, as is well
known \cite{Hamann:2008bc}. The latter ratio is, in
turn, a sensitive function of the $\nu_e$ and $\bar{\nu}_e$ energy
distribution functions, and these can be affected by BSM issues like
lepton numbers, flavor mixing, sterile neutrino states, heavy particle
decay, etc.

Handling and analyzing the observed data while imposing
self-consistency over multiple epochs in the early universe can
require new procedures. The approach developed in
Ref.\ \cite{Hamann:2008bc} and employed in Planck XVI
\cite{PlanckXVI:2013} has been highly successful.  There,
self-consistency is imposed approximately by the addition of a term in
the log-likelihood function. A very small theoretical uncertainty
ensures the posterior distributions are close to the corresponding
theoretical values.  However, when we impose self-consistency among
the different epochs the need may arise to solve for the various
derived parameters iteratively.  Returning to the helium/\neff\
example discussed above, $Y_P$ and the recombination history of the
universe both depend on the radiation energy density, usually
parametrized by \neff. Subsequently, the recombination history is
affected by the the primordial abundance of helium. The values of
these two derived parameters, in turn, affect CMB observables, which
recommends an iterative and therefore self-consistent approach.

Ideally, we would want a procedure to self-consistently treat neutrino
and BSM physics from the post-QCD epoch to the onset of
non-linearities in large scale structure (LSS).
Recently, the quantum-kinetic equations (QKEs)
governing neutrino flavor evolution in dense environments
have been derived from first principles
\cite{VFC:QKE, 2013PhRvD..87k3010V, 2014PhRvD..90l5040S}.
Such a program to treat neutrino physics in the early universe would
incorporate a QKE treatment
through weak decoupling at the very least.
Weak freeze-out and BBN necessitate a neutrino-energy Boltzmann-equation or
QKE treatment fully coupled to a nuclear reaction network. The
recombination, photon decoupling, and advent of LSS epochs require a
Boltzmann-equation treatment of neutrino clustering.  Verification of any models
related to neutrinos and BSM physics would then rely on agreement with
direct cosmological observables, including but not limited to:
primordial abundances; CMB power spectra; and the total matter power
spectrum.

As outlined, this is a challenging undertaking. We therefore employ a
limited approach to show why self-consistency is required and how such
a treatment can be efficacious in probing some issues in neutrino
sector physics.  To that end, we calculate the ratio of sound horizon
to photon diffusion length, $r_s/r_d$, as a simple parametrization of
the CMB, as described in detail in Sec.\ \ref{sec:method}.  It would, of
course, be preferable to compute the full CMB power spectra
in the self-consistent manner
described above. There are two limiting factors that temper this
ambitious proposal, however. The first is that the observable
$r_s/r_d$ is largely insensitive to the poorly constrained equation of
state of the dark, vacuum energy component while the CMB power spectra
are not.  Additionally, current computations of the CMB power spectra
\cite{Lewis:1999bs} would need to be generalized to the BSM scenarios we
contemplate here.

Our procedure for calculating derived cosmological quantities utilizes
the neutrino occupation probabilities. Since we are developing the
capacity to compute the effects of BSM physics that couples to the
active neutrino sector, we do not restrict the form of the neutrino
distribution functions to that of equilibrium Fermi-Dirac
distributions. General forms are permitted and may be handled
analytically or numerically. Neutrino occupation probabilities are
taken in the present work in the flavor eigenbasis during weak
decoupling, weak freeze-out and BBN.\footnote{This is a matter that is
properly resolved by utilizing a QKE approach \cite{VFC:QKE}, the
subject of a future study.} During recombination and last scattering,
we transform the occupation probabilities to the mass eigenbasis. When
we consider the case of massive neutrinos, the statistic for the sum
of the light neutrino masses, \summnu, is simply the sum of the three
lightest mass eigenstates.

We might further clarify the fact that this limited approach does not
rely on the assumption that the radiation energy density can be
described in terms of the single parameter \neff. Indeed, an original
motivation for using $r_s/r_d$ as a proxy for \neff\ (see
Sec.\ \ref{sec:method}) is the fact that the usual definition for \neff\
does not apply to non-equilibrium neutrino distributions. A corollary
of this observation is the possibility that \neff\ may depend
on the scale factor $a(t)$; that is, we do not assume that \neff\ is
independent of scale factor. When extended, our
approach is able to handle BSM scenarios such as the decay of massive
sterile neutrinos and other weakly interacting massive particles,
which result in neutrino distributions that may differ substantially
from a Fermi-Dirac distribution \cite{FKK:2011di}.

Our approach aims to be general; it does not rely on particular
cosmological models but is appropriate to the class of $\Lambda$
cold-dark matter (\lcdm) models. It extends and explicates recent work
\cite{Hamann:2008bc,Shimon:2010cb,Nollett:2011ho,Hou:2013da} on the
consistent incorporation of precision observations of the CMB
\cite{PlanckXVI:2013} and observations of the light element
abundances of helium and deuterium (mass fraction $Y_P$ and relative
abundance D/H).  The deuterium abundance
\cite{Pettini:2012yd,Cooke:2014do} is now more precisely measured by a
significant factor (4 or 5) in relative precision than $Y_P$. Our
results indicate that deuterium is sufficiently well determined to be
incorporated into CSM analyses as a fixed prior. 

The present work is a first result in an ongoing campaign to
incorporate neutrino energy transport into a self-consistent treatment
of BBN and recombination. Our approach is being implemented in {\sc
Fortran90/95} as a suite of codes under the working title of
``BBN Unitary Recombination Self-consistent Transport''
({\sc burst}), which will be made publicly available for use on parallel
computing platforms using {\sc openmpi}. The current work is a
prerequisite in an ongoing collaboration to develop codes that
consistently handle BBN and neutrino energy transport from weak
decoupling to the advent of LSS.

The paper is outlined as follows.  In the following section,
Sec.\ \ref{sec:bckg}, we discuss our treatment of BBN and relevant CMB
physics.  Section \ref{sec:method} explains the self-consistent,
iterative approach to determining \neff\ using $r_s/r_d$.  Section
\ref{sec:4pts} employs a dark-radiation model to test the accuracy of
predictions made by {\sc burst} against observations.  Section
\ref{sec:bsm} investigates a number of `test problems' from BSM physics
suited for {\sc burst}. We conclude in Sec.\ \ref{sec:outl} with future
applications of the results given in this work.  Throughout this
paper, we use natural units where $\hbar=c=k_B=1$.

\section{Background}\label{sec:bckg}
\subsection{Big-bang nucleosynthesis}\label{ssec:bbn}
Recently there has been a dramatic increase in precision in the
determination of D/H \cite{Cooke:2014do}.
This improvement in observational precision
drives the need to improve the standard tools used to calculate
observables during the BBN and CMB epochs. This development allows
us to consider improved descriptions of nuclear physics and
non-standard particle and cosmological models. 

We have developed a generalized BBN subroutine, part of the larger
{\sc burst} code.  The current BBN routine handles general
distribution functions for particles out of equilibrium.  It
numerically integrates the binned neutrino occupation probabilities to
obtain thermodynamic variables and number and energy densities for
$\nu_e, \bar{\nu}_e, \nu_\mu, \bar{\nu}_\mu, \nu_\tau,
\bar{\nu}_\tau$.  It is worthwhile, therefore, to summarize our
approach to BBN even though it is substantially similar in some
respects to that given in Refs.\
\cite{Wagoner:1966pv,Wagoner:1969sy,SMK:1993bb}. 

We assume that the cosmic fluid is homogeneous and isotropic and that
the comoving number density of baryons is covariantly conserved.
Neutrinos can experience, in addition to gravity, essentially
arbitrary interactions within the Boltzmann transport
approximation.\footnote{However, the code can handle limited
generalizations of the Boltzmann approximation to incorporate effects
associated with neutrino quantum kinetics.} We allow for the
possibility that neutrinos may decouple from the plasma with
non-equilibrium distributions. This assumption implies that there may
be deviations from the Fermi-Dirac momentum distribution
\cite{Dolgov:1997ne,Gnedin:1998ne,Mangano:3.046}. In fact, it was this
observation that provided initial motivation for developing the
current approach.

BBN codes evolve light nuclide abundances $Y_i(t)$, defined as
\begin{align}
   \label{eqn:yidef}
   Y_i(t) &= \frac{n_i(t)}{n_b(t)},
\end{align}
as a function of comoving time $t$ in the background of the \flrw\
(FLRW) geometry.  Here $n_i(t)$ is the proper number density of nuclide
$i=n$, $p$, $^2\mbox{H}$, $^3\mbox{He}$, $^4\mbox{He},\ldots$ and $n_b(t)$ is the
proper baryon number density.
We will focus on two specific nuclides in this paper: the $^4\mbox{He}$
mass fraction ($Y_P\equiv4Y_{^4He}$),
and the $^2\mbox{H}$ relative abundance ($\mbox{D/H}\equiv Y_D/Y_H$).
The evolution starts from a time when the plasma temperature $T$ is
near 30 MeV.  Weak equilibrium obtains at this temperature.  The time
dependence of the metric is determined by the energy density $\rho(a)$
as 
\begin{align}
   \label{eqn:fe1}
   H(a) &= \frac{\dot a}{a} = \sqrt{\frac{8\pi}{3m_{\rm pl}^2}\rho(a)},
\end{align}
where $H(a)$ is the Hubble parameter, $m_{\rm pl}$ is the Planck mass, and the `dot' indicates
differentiation with respect to the FLRW-coordinate time $t$.  The energy density
is computed as the integral of the single-particle energy over the
momentum distribution:
\begin{align}
   \label{eqn:rhof}
   \rho(a) &= \sum_j \int\frac{d^3p}{(2\pi)^3} f_j(p;a)
   \sqrt{p^2 + m_j^2}
\end{align}
where the sum over $j$ reflects all species contributing to the energy density,
including but not limited to: photons, baryons, dark matter, $e^\pm$, and neutrinos. 
Consistent with the above discussion, we do not need to assume a well
defined temperature for any of the cosmic species. The time
dependence of the distribution function $f_i(p;a)$ is indicated by the
presence of the scale factor $a(t)$ in its argument. Evolution in
time of the distribution functions is accomplished by solving
transport equations, such as the Boltzmann equation,
in FLRW geometry. 

Given initial conditions for the temperature $T$ and momentum
distributions of the cosmological constituents (assumed to be in
equilibrium through weak and strong interactions), the BBN code
determines the evolution, with respect to the scale factor $a(t)$
(related to time as $dt = da/(aH(a))$), of the temperature of the
plasma $T(a)$, the electron chemical potential $\mu_e(a)$, and the
nuclide abundances relative to baryon number $Y_i(a)$.

The interactions of the light nuclear species is governed by the
nuclear reaction network.\footnote{The nuclear reaction network should
   be derived from reaction cross sections that are governed by the
   principles of quantum mechanics, such as unitarity. We are
currently developing a unitary reaction network for application in
future work.} The reaction network, which is determined by a chosen
set of nuclides and the thermally averaged reactivities
$n_{\alpha_1}n_{\alpha_2}\langle \sigma_{\beta\alpha}
v_\alpha\rangle$, is proportional to the rate of change of the number
density of nuclides participating either in the initial state
$n_{\alpha_i}$ or final state $n_{\beta_j}$ where $i,j$ indexes
particles (up to 3) in the initial $\alpha$ or final $\beta$ channel
for the process $\alpha\to\beta$; that is, $\alpha$ and $\beta$ are
two- or three-body reaction channels and $\alpha_i$ is the
$i^{\mbox{\scriptsize th}}$ nuclide of the channel $\alpha$. 
The nuclear reaction network includes nuclides with mass number $A \le
9$.  Our code allows for the inclusion of additional nuclides with
$A>9$, but we maintain the smaller network as the larger network
provides no new insights for this paper.  The nuclides are taken to be
in thermal equilibrium with the photon--electron plasma. We are
currently employing an updated \cite{Fuller:2010nn} version of the
reaction network from Ref.\ \cite{SMK:1993bb}.  We also couple in all
relevant $e^\pm$ and neutrino-induced weak interactions (charged and
neutral current) \cite{2010PhRvD..82l5017F}.

\subsection{Cosmic Microwave Background}\label{ssec:cmb}
The physics of BBN and photon decoupling are related
through the epoch of recombination by several mechanisms including,
importantly, He {\sc i} and He {\sc ii} recombination. The
recombination rates depend sensitively on the amount of \jHeiv\
created in BBN \cite{Trotta:2004bc,Ichikawa:2006bc,Hamann:2008bc}.
The \jHeiv\ dependence also determines the photon diffusion damping
wave number $k_d$ or, equivalently, the photon diffusion length,
defined as $r_d \equiv\pi/k_d$.  This quantity depends strongly on the
recombination history through the free-electron fraction, $X_e(a)$,
which, in turn, depends on the helium abundance (see
Sec.\ \ref{sssec:dl} below). 

We relate the diffusion length directly to an observed angular size
$\theta_d$ through the angular diameter distance $D_A$ at the
epoch of last scattering (as described in detail below). Since
$D_A$ depends on the vacuum (dark) energy equation of state, which is
poorly understood, we employ the ratio $\theta_s/\theta_d$, equivalent to $r_s/r_d$
(see Eqs.\ \eqref{eqn:ths} and \eqref{eqn:ddl}),
where $\theta_s$ is the angle subtended by the sound horizon, which eliminates explicit
dependence on the dark energy equation of state.

Modern computations of the recombination history of the early universe
are available in accurate and well tested codes such as {\sc
HyRec} \cite{AliHamoud:2011hr}, {\sc CosmoRec} \cite{Chluba:2011cr}, and
the fast, approximate computations given in Ref.\cite{SSS:2000rc} by
the code {\sc RecFast} \cite{SSS:1999nc}.
Our
recombination history compares well with {\sc RecFast}, agreeing at
the level of $<3\%$ over the recombination epoch. 
The present accuracy suffices for the
purposes of the current exploratory work.

\subsubsection{Sound horizon}
\label{sssec:sh}
The sound horizon
is defined as
\begin{align}
\label{eqn:shdef}
r_s (a_{\gamma d}) &=
\int_0^{a_{\gamma d}} da\, \frac{1}{a^2 H(a) \sqrt{3( 1+R(a))}}
\end{align}
where $a_{\gamma d}$ is the scale factor at photon
decoupling.\footnote{Reference \cite{PlanckXVI:2013} takes $a_{\gamma
d}\to a_\star$ where the optical depth is unity.} 
The ratio $R(a)$ is given as
\begin{align}
\label{eqn:Rdef}
R(a) &= \frac{3\rho_b(a)}{4\rho_\gamma(a)},
\end{align}
where $\rho_b$ and $\rho_\gamma$ are the baryon rest mass
and photon field energy densities, respectively.

The angular size of the sound horizon $\theta_s$\footnote{Reference
\cite{PlanckXVI:2013} takes $\theta_s \to \theta_\star$.},
determined from the spacing of the acoustic peaks in the CMB
temperature power spectrum, is related to the sound horizon by the
angular diameter distance $D_A(a_{\gamma d})$ at photon decoupling
(scale factor $a_{\gamma d}$) as
\begin{align}
   \label{eqn:ths}
   \theta_s(a_{\gamma d}) 
   &= a_{\gamma d}\frac{r_s(a_{\gamma d})}{D_A(a_{\gamma d})},
\end{align}
since the angle $\theta_s$ is small. The angular diameter distance is
\begin{align}
   \label{eqn:add}
   D_A(a) &= a\int_a^{a_0} da'\, [a'^2 H(a')]^{-1}.
\end{align}
The quantity $D_A(a_{\gamma d})$ depends on the vacuum (dark)
energy equation of state, which is not very well understood. Our
approach will eliminate dependence on this poorly constrained
component of the energy density.

\subsubsection{Free-electron fraction} \label{sssec:reco} 
We have written an independent code to calculate the free-electron
fraction, $X_e$.  The results agree well with Ref.\ \cite{SSS:1999nc}
({\sc RecFast}). In fact, the agreement is within $2\%$ for most of
the range $10^{-4} \lesssim a/a_0 \lesssim 10^{-3}$ ($10^4 \gtrsim z
\gtrsim 10^3$).  
The code allows significant deviations from the model parameters of
\lcdm; it should be used with caution, however, since the effective
three-level treatments for helium and hydrogen recombination have been
optimized for near-standard model parameters.

The number density of free and total electrons is
\begin{align}
   n_e^{\rm (free)} &= n_p + n_{\heii} + 2n_{\heiii},\\
   \label{eqn:netot_def}
   n_e^{\rm (tot)} &= n_b\left(1 - \frac{Y_P}{2}\right),
\end{align}
where $n_b$, $n_p$, $n_{\heii}$, $n_{\heiii}$ are the proper number densities
for baryons, protons, singly- and doubly-ionized helium, respectively.
We write the free-electron fraction as
\begin{align}
   \label{eqn:xe_def}
   X_e \equiv \frac{n_e^{\rm (free)}}{n_e^{\rm (tot)}} 
   \equiv X_p + X_{\heii} + 2X_{\heiii},
\end{align}
so defined to take values in the range $0\le X_e \le 1$.

We follow Refs.\ \cite{1968ApJ...153....1P,1968ZhETF..55..278Z} and
consider the simplification of the multi-level hydrogen and helium
atoms to that of an effective three-level system which includes the
ground $n=1$ state, the first excited $n=2$ states, and the continuum.
All other excited states are assumed to be in equilibrium with the
$2s$ state. We treat He {\sc ii} recombination approximately
\cite{SSS:2000rc} (via the Saha equation) since it is essentially
complete at the advent of the epoch of He {\sc i} recombination.  The
He {\sc iii} contribution, therefore, in the Boltzmann equation for He
{\sc ii} is negligible.

Boltzmann equations for H {\sc ii} and He {\sc ii} contain a
thermally-averaged cross section and relative velocity $\langle\sigma
v\rangle$.
We use Case B recombination coefficients
for the $\langle\sigma v\rangle$ of both neutral hydrogen
\cite{1991A&A...251..680P} and helium \cite{1998MNRAS.297.1073H}.
Along with a Saha equation for He {\sc iii}, the Boltzmann equations
for H {\sc ii} and He {\sc ii} are a coupled set of ordinary
differential equations constituting a recombination network to model
the ionization history of the universe prior to photon decoupling.

\subsubsection{Photon diffusion damping length}\label{sssec:dl}
In the tightly coupled limit, photon diffusion damping is characterized through the damping wave
number \cite{Silk:1968dd,Zaldarriaga:1995dd,Hu:1997dd,Weinberg:2008co}
\begin{align}
   \label{eqn:dl}
   k_d^{-2} &= 
   \int_0^{a_{\gamma d}} \frac{da}{a^2 H(a)}\, 
   \frac{1}{an_e(a)\sigma_T}
   \frac{
   R^2(a)+\frac{16}{15}(1+R(a))
   }{
   6(1+R(a))^2},
\end{align}
where $\sigma_T$ is the Thomson cross
section. Here, we have assumed that moments of the temperature
fluctuation higher than the quadrupole make a negligible contribution in
the linearized Boltzmann equation for the photon distribution.

Equation \eqref{eqn:dl} requires the free-electron fraction, $n_e(a)$,
determined in the recombination history, as discussed in the previous
section. The free-electron fraction, over the course of the
recombination history, depends strongly on the primordial helium mass
fraction $Y_P$.  The fraction $Y_P$, in turn, depends on the
relativistic energy density, which is often parametrized in terms of
the parameter \neff:
\begin{align}
\label{eqn:rhor-neff}
\rho_r &=
2\left[ 1 + \frac{7}{8}\left(\frac{4}{11}\right)^{4/3}\neff\right]
\frac{\pi^2}{30}T^4.
\end{align}
Equation \eqref{eqn:rhor-neff} is only valid after the epoch of
$e^\pm$ annihilation. The point that BBN and recombination are related
is well known \cite{Hamann:2008bc} but has not been implemented
self-consistently as a constraint for general, BSM physics model
cosmologies. We return to it after describing the relation between
$r_s$ and $k_d$ to directly observable quantities given by the CMB
power spectrum.

The observed diffusion angle $\theta_d(a_{\gamma d})$
\cite{PlanckXVI:2013} is related to the diffusion damping length, $r_d
= \pi/k_d$ as
\begin{align}
   \label{eqn:ddl}
   \theta_d(a_{\gamma d}) &= a_{\gamma d}
   \frac{r_d(a_{\gamma d})}{D_A(a_{\gamma d})},
\end{align}
with the same stipulation regarding the smallness of the angle in Eq.\eqref{eqn:ths}.

\section{$r_s/r_d$ as a proxy for \neff}\label{sec:method} In this
section we describe our method for determining \neff\ in detail. We
introduce two variants of \neff. When referring to the radiation
energy density equation \eqref{eqn:rhor-neff}, which takes as an input
the quantity \neff, we designate \neff\ as \neffth. When considering
general cosmologies, perhaps with BSM physics, we deduce the value of
\neff\ from the observable quantity $r_s/r_d=\theta_s/\theta_d$, described in this
section, and designate it as \neffobs. The simplest cosmologies for
which Eq.\ \eqref{eqn:rhor-neff} obtains, having negligible neutrino
mass, standard model constituents and no energy transfer between
species have \neffth=\neffobs.

We consider a test input cosmology that is non-standard yet
substantively similar to \lcdm.
We proceed by determining $Y_P$ at temperature $T\sim0.1$ MeV using
the BBN network of {\sc burst} \cite{Fuller:2010nn}.  The principal
observational cosmological input at this time is \barnum. Also input
and incorporated into the BBN subroutine is any neutrino and BSM
physics that constitute the test cosmology. Subsequently, we compute
the recombination history of the universe from early times
($a/a_0\sim10^{-7}$) to the current epoch.  Specific observational
inputs include \barnum, $\omega_c$ (where for cold dark matter
$\omega_c\equiv\Omega_c h^2$), $Y_P$, and $H_0$ (the Hubble constant,
$H_0 = H(a_0)$).  For the purposes of the present discussion, other
inputs of particular importance include neutrino occupation
probabilities (as output from a neutrino transport calculation of weak
decoupling that is fully coupled to BBN) and neutrino rest masses.
The neutrino energy density of the neutrino seas is calculated by
writing the occupation probabilities in the mass eigenbasis
\cite{KF:2008PRL}. We emphasize the fact that \neffth is {\em not}
input as a base parameter; this is of paramount import in the
present approach. The recombination history, $n_e(a)$ determines the
optical depth as a function of scale factor:
\begin{align}\label{eqn:tau}
  \tau(a)\equiv
  \int^{a_0}_{a}
  \frac{da^{\prime}}{a^{\prime2}}\,a^{\prime}n_e(a')\sigma_T
\end{align}
We define the scale factor at photon decoupling $a_{\gamma d}$ such
that $\tau(a_{\gamma d})\equiv1$.
In this definition, we do not include the effects of cosmic reionization
when calculating $n_e(a)$ for use in Eq.\ \eqref{eqn:tau} \cite{PlanckXVI:2013}.
We apply $a_{\gamma d}$ and the input cosmology to equations \eqref{eqn:shdef}
and \eqref{eqn:dl} to compute the sound
horizon and photon diffusion length, respectively.
We arrive in this
way at a ratio $(r_s/r_d)^{\rm (inp)}$ for our input cosmology.

\begin{figure}[ht]
   \begin{center}
   \includegraphics[scale=0.6]{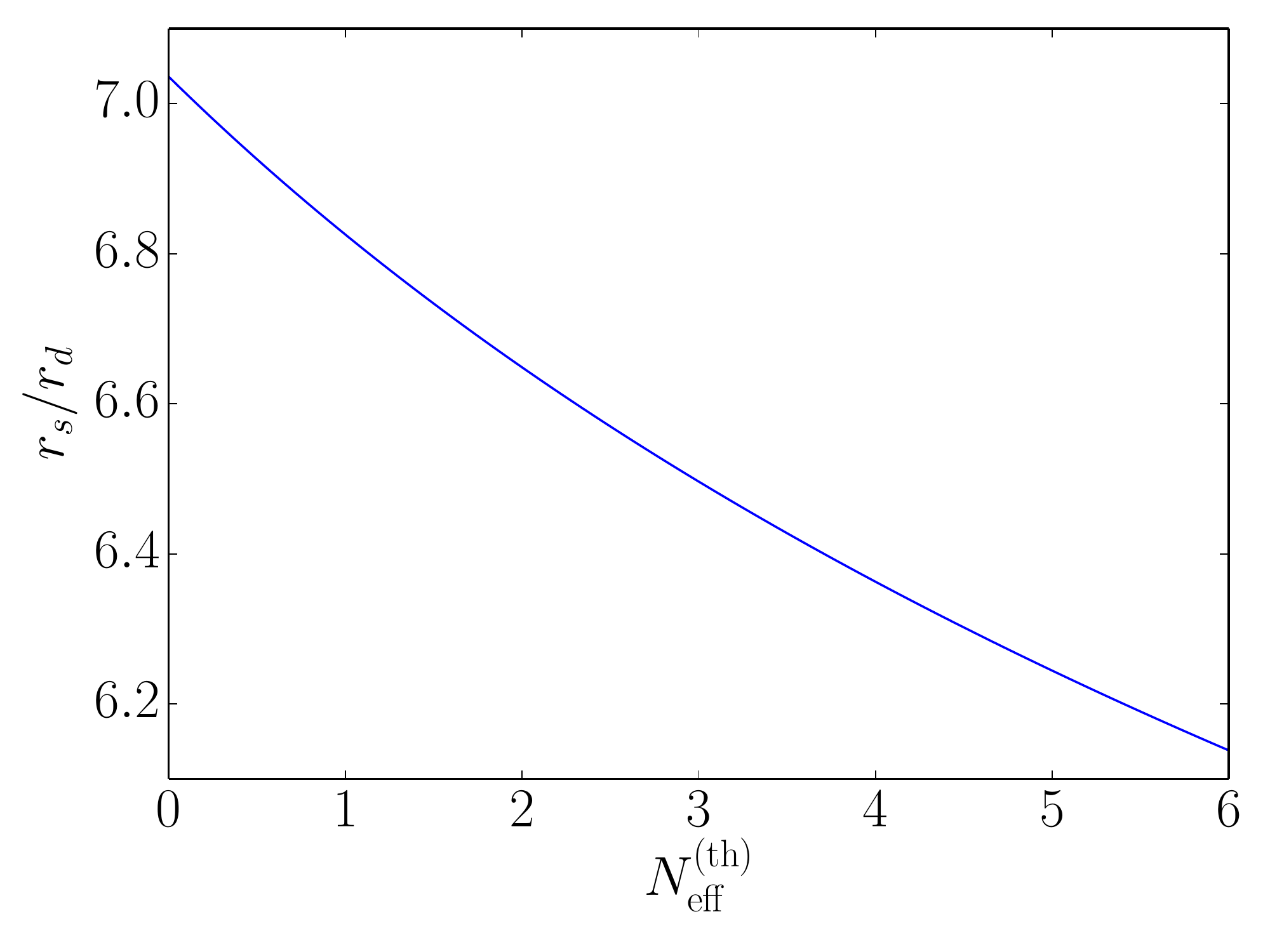}
   \caption{\label{fig:rrvneff}Ratio of the comoving coordinate of the
      sound horizon radius $r_s$ to that of the photon diffusion
      length $r_d$ as a function of \neffth for cosmological parameter
      values $Y_P=0.2425$, $\barnum=0.22068$, and $\omega_c=0.12029$.}
   \end{center}
\end{figure}

Our immediate objective is to determine a value of \neff (here
termed \neffobs) corresponding to this value for $(r_s/r_d)^{\rm
(inp)}$.  We map out a range of values of $r_s/r_d$ that correspond
to the same input cosmology as that used in calculating
$(r_s/r_d)^{\rm (inp)}$, with one significant difference.  We
parametrize all of the neutrino and BSM physics into the single
\neffth parameter.  We then use \neffth to calculate the radiation energy
density in Eq.\ \eqref{eqn:rhor-neff} to determine the Hubble
rate. We vary \neffth to compute the function
\begin{align}\label{eqn:rsrdfunc} 
r_s/r_d = r_s/r_d[\barnum, \omega_c, Y_P,\ldots;\neffth],
\end{align} 
shown in Fig.\ \ref{fig:rrvneff}. Since $r_s/r_d$ is a one-to-one
function of \neffth, we may invert Eq.\ \eqref{eqn:rsrdfunc} to obtain
$\neffth = \neffth[r_s/r_d]$. The final step is to evaluate the
previous function with our input cosmology ratio, {\em i.e.}\
$\neffobs = \neffth[r_s/r_d = (r_s/r_d)^{\rm (inp)}]$, to obtain a
value of \neffobs.
As an example, we take the best-fit values from Ref.\ \cite{PlanckXVI:2013}
combined with WMAP Polarization data
($100\theta_s=1.04136$ \& $100\theta_d=0.161375$) to obtain
$(r_s/r_d)^{\rm (inp)} = 100\theta_s/100\theta_d = 6.45304$.  This corresponds to a value
$\neffobs=3.31$ on Fig.\ \ref{fig:rrvneff}, in line with the best-fit
value $\neff=3.25$ ($3.51^{+0.80}_{-0.74}$ at 95\% limits) \cite{PlanckXVI:2013}.
We again note that for the simplest cosmologies the
two generally distinct functions $\neffth[r_s/r_d]$ and
$\neffobs[r_s/r_d]$ reduce to the same function and have
$\neffth=\neffobs$.

Figure \ref{fig:rrvneff}, which shows the function $r_s/r_d[\neffth]$,
demonstrates an important constraint between phenomena occurring during
the epochs of BBN and recombination/photon decoupling.  For a given
input cosmology ($\barnum, \omega_c, \ldots$), the graph of $r_s/r_d$
as a function of \neffth requires a computation of $n_e(a)$ to obtain
$r_d$ for each value of \neffth.

We can understand Fig.\ \ref{fig:rrvneff} qualitatively by a simple
scaling argument. We expect that,
as the radiation energy density increases with increasing \neffth, the
sound horizon and the diffusion length will decrease with the
increasing Hubble rate $H(a)$. The sound horizon decreases due to the
increased energy density driving a more rapid expansion and a decrease
in the sound speed, $c_s=[3(1+R(a))]^{-1/2}$. The diffusion length
increases, naively, due to a decrease in the scattering rate driven by
the reduced Hubble time. Caution should be taken when using such naive
scaling arguments. For example, the non-trivial dependence of the 
recombination history leads to counterintuitive effects in the
\neffobs dependence on \summnu \cite{GFP-PRL:2014mn}.
Ref.\ \cite{GFP-PRL:2014mn} demonstrates that where a naive scaling
argument would suggest an increase in \neffobs with increasing \summnu, the
non-trivial dependence of the recombination history on \summnu implies
\neffobs decreases monotonically and rapidly with increasing \summnu
(See Sec.\ \ref{ssec:summnu}).

\section{Verification of cosmological parameters with dark radiation}
\label{sec:4pts}
We adopt the point of view advocated in Refs.\
\cite{Trotta:2004bc,Ichikawa:2006bc,Hamann:2008bc} that the constraint
provided by predictions of BBN should be incorporated simultaneously
with constraints due to recombination effects in the extraction of
cosmological parameters. We also require, as previously discussed,
that BBN and recombination be solved iteratively. In this section, we
demonstrate the self-consistent extraction of these parameters by
employing a simple model for the radiation energy density that avoids
solving, for example, the Boltzmann equation, for the set of
distribution functions of the cosmic constituents, in particular the
neutrinos.

The model we explore in this section is identical to \lcdm\ except for
one additional constituent: dark radiation.  The dark radiation energy
density is radiation at all epochs and does not interact with the
other energy-density constituents through any force except
gravitation.

The total energy density as given in terms of radiation, matter, and
vacuum energy components is
\begin{align}
   \label{eqn:rho-tot}
   \rho(a) &= \rho_r(a) + \rho_m(a) + \rho_v(a),
\end{align}
which depend on the scale factor $a$ as $a^{-4}$, $a^{-3}$, $a^0$,
respectively, as long as there is no
energy transfer between the species.  We assume for the purposes of
this section that the neutrinos are massless, always
acting as radiation energy density.
The matter energy density consists of contributions from baryons and
cold dark matter. Modeling the matter as a pressureless gas and observing that
the comoving matter energy density is conserved, we write the proper 
energy density as
\begin{align}
   \rho_m = \frac{3H_0^2\,m_\text{pl}^2}{8\pi}(\Omega_b +
   \Omega_c)\left(\frac{a_0}{a}\right)^3.
\end{align}

The vacuum energy is the least understood of the energy densities.
Assuming the universe to be critically closed, the sum of the
three energy densities must be equal to the critical energy density,
specified only by the Hubble rate at the current epoch.  Hence,
$\rho_v = \rho_c - \rho_{r,0} - \rho_{m,0}$.  The vacuum energy
density is negligible at all epochs of interest in this paper but is
included for completeness. 

The radiation component is given as:
\begin{align}
   \rho_r = \rho_\gamma + \rho_\nu + \rho_\text{dr}
\end{align}
where $\rho_\gamma$ is the photon energy density,
$\rho_\nu$ is the neutrino energy density, and
$\rho_\text{dr}$ is the dark-radiation energy density.  We parametrize
$\rho_\text{dr}$ as
\begin{align}
   \rho_\text{dr} =
   \frac{7}{4}\left(\frac{4}{11}\right)^{4/3}
   \frac{\pi^2}{30}T^4\,\delta_\text{dr},
\end{align}
where $\delta_\text{dr}$ is the dark-radiation parameter and is always
assumed to be non-negative.
In principle, we could entertain negative
values of $\delta_\text{dr}$ since it is an adjustable parameter of
$\rho_r$.  Such a change requires a fundamental reworking of the
\lcdm\ model so that $\sum_i \Omega_i = 1$. These non-standard
cosmologies obtain when considering, for example,
neutrino oscillations. These models, however, are not continuously
connected with our model at $\delta_\text{dr} = 0$, for any values of
the parameters, thus motivating the maintenance of $\delta_\text{dr}>0$.

We write $\neffth = 3 + \Delta\neffth$ and assume that the
contribution to \neffth from $\rho_\nu$ is $3$. Then the contribution
from $\rho_\text{dr}$ is given as $\Delta\neffth$.  We see then that
$\Delta\neffth = \delta_\text{dr}$.  This is simply a restatement of
the fact from the previous section, Sec.\ \ref{sec:method}, that
$\neffobs=3+\Delta\neffth=\neffth$ for `standard' cosmologies.  It
is, therefore, unnecessary for this simple dark-radiation model, to
deduce \neffobs from $r_s/r_d$ since \neffth=\neffobs by construction.
This model of dark radiation is the usual model applied, for example in
Ref.\ \cite{PlanckXVI:2013} and we explore, in this section, the
predictions of the present {\sc burst} code to verify our results
against those of prior results within the community. 
We will use \neffobs, for the remainder of this section, to denote the
``effective number of relativistic degrees of freedom\footnote{We refer
to \neff\ as the effective number of relativistic degrees of freedom
although there are factors that complicate this interpretation, among them
the temperature parameter, and fermionic nature of neutrinos.}.'' 

\begin{figure}[h]
   \begin{center}
      \includegraphics[scale=0.60]{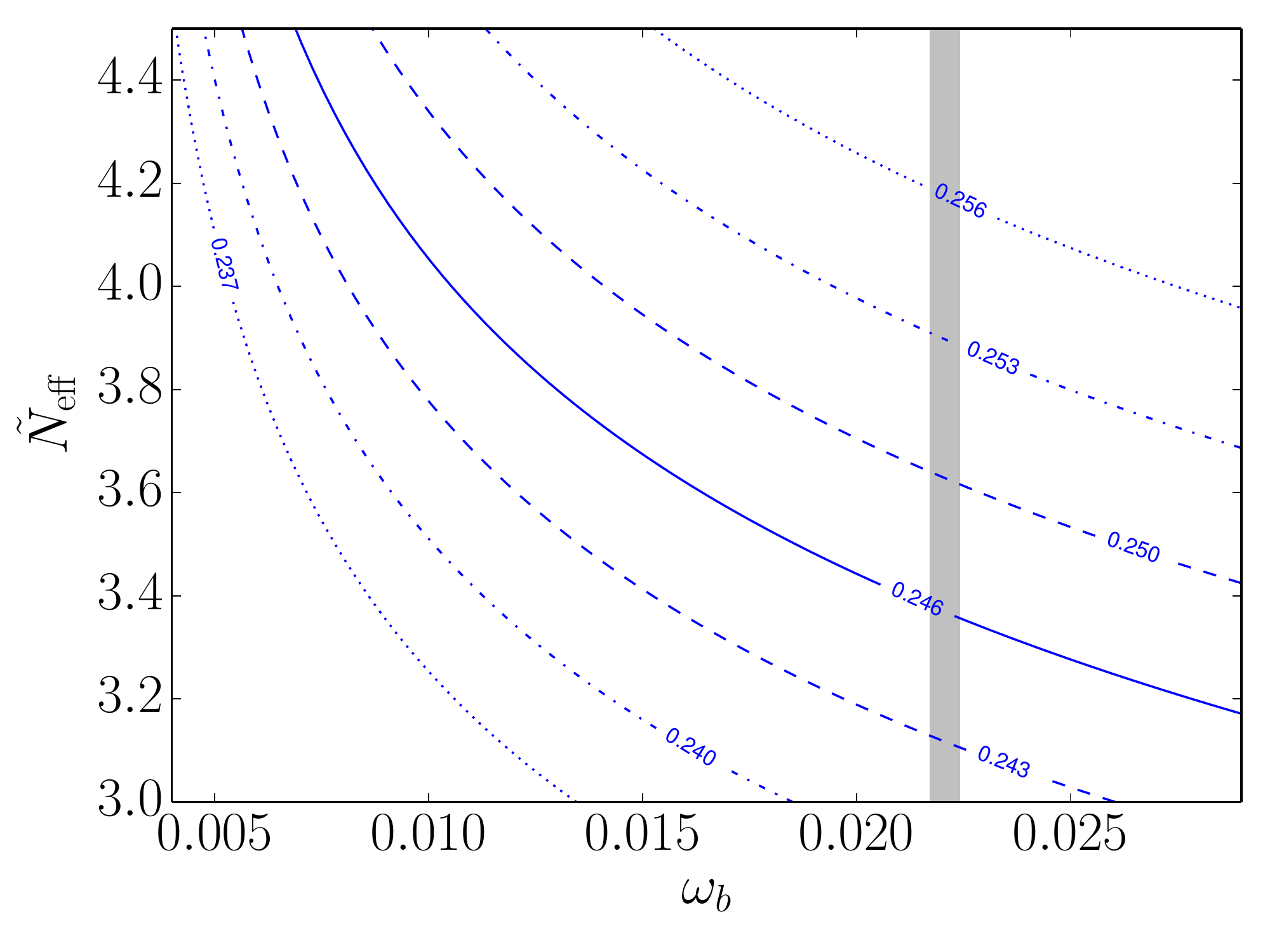}
      \includegraphics[scale=0.60]{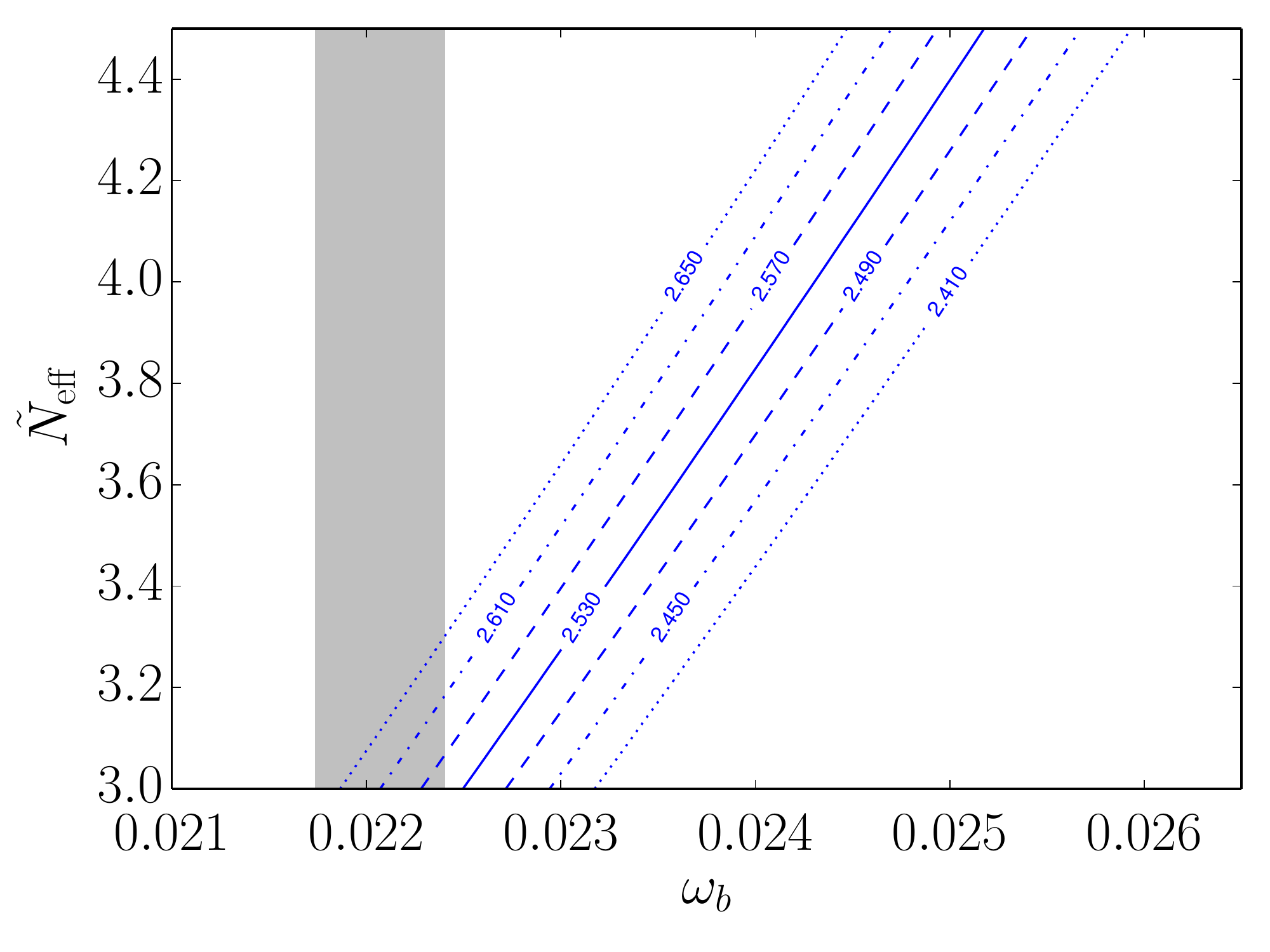}
   \end{center}
   \caption{\label{fig:nvocy}(Top) \neffobs\ plotted against \barnum
      for contours of constant values of $Y_P$ (labeled by mass
      fraction).  The solid curve is the preferred value of Ref.\
      \cite{Aver:2013ue}.  The contours are spaced by $\Delta
      Y_P\approx0.003$.  (Bottom) \neffobs\ versus \barnum for
      contours of constant values of $10^5\times\mbox{D/H}$.  The
      solid curve is the preferred value of Ref.\ \cite{Cooke:2014do}.
      The contours are spaced by $\Delta(10^5\times\mbox{D/H})=0.04$.
   In each case, abundances are determined in a self-consistent BBN
calculation.}
\end{figure}

\begin{figure}
   \begin{center}
      \includegraphics[scale=0.60]{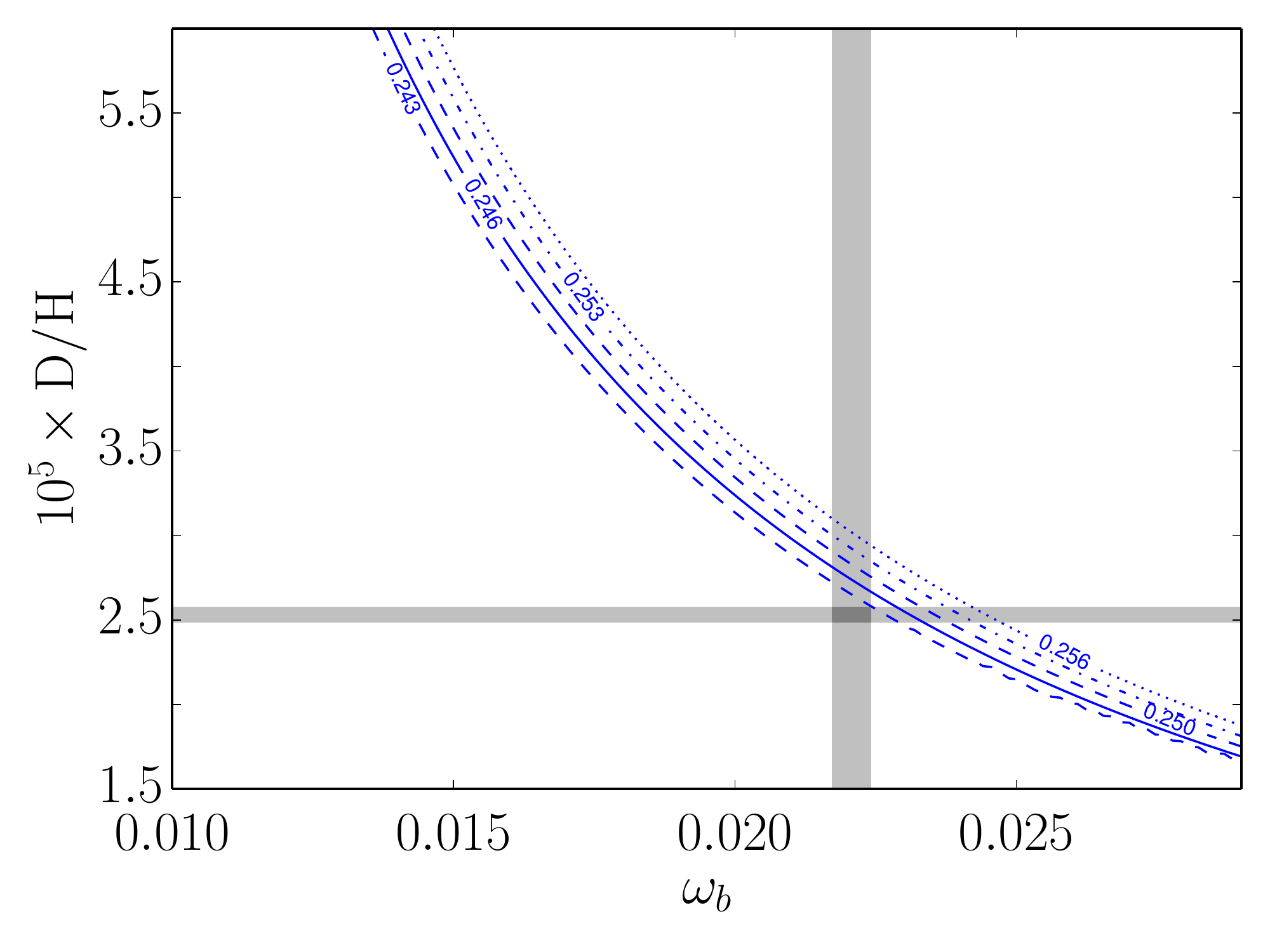}
      \includegraphics[scale=0.60]{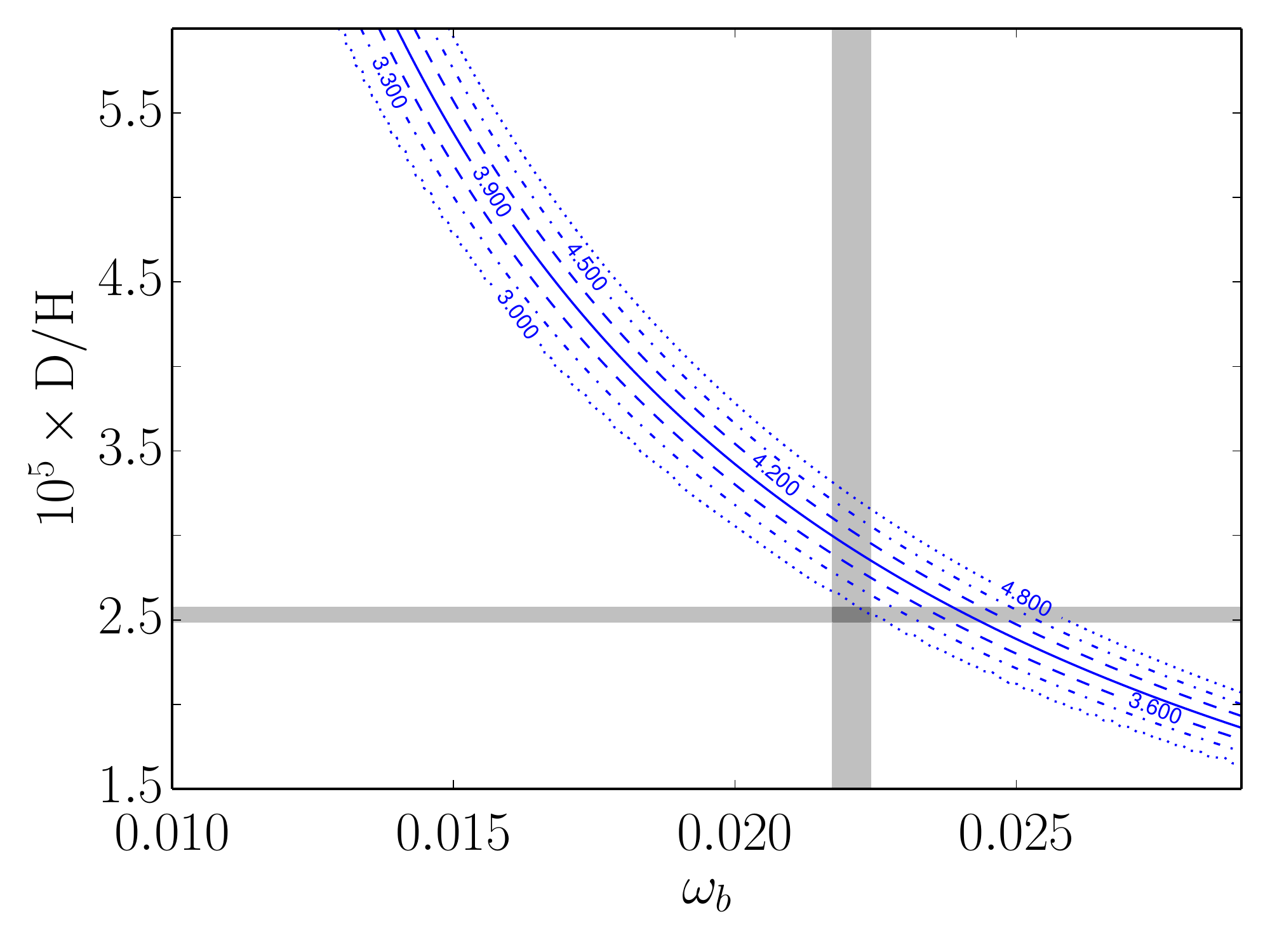}
   \end{center}
   \caption{\label{fig:DHvbn}(Top) $10^5\times\mbox{D/H}$
   plotted against \barnum for contours of constant $Y_P$.
   The solid curve is the preferred value of Ref.\ \cite{Aver:2013ue}.
   The contours are spaced by $\Delta Y_P\approx0.003$.
   (Bottom) $10^5\times\mbox{D/H}$ versus \barnum
   for contours of constant \neffobs.
   The contours are spaced by $\Delta\neffobs=0.3$}
\end{figure}

\begin{figure}
   \begin{center}
      \includegraphics[scale=0.60]{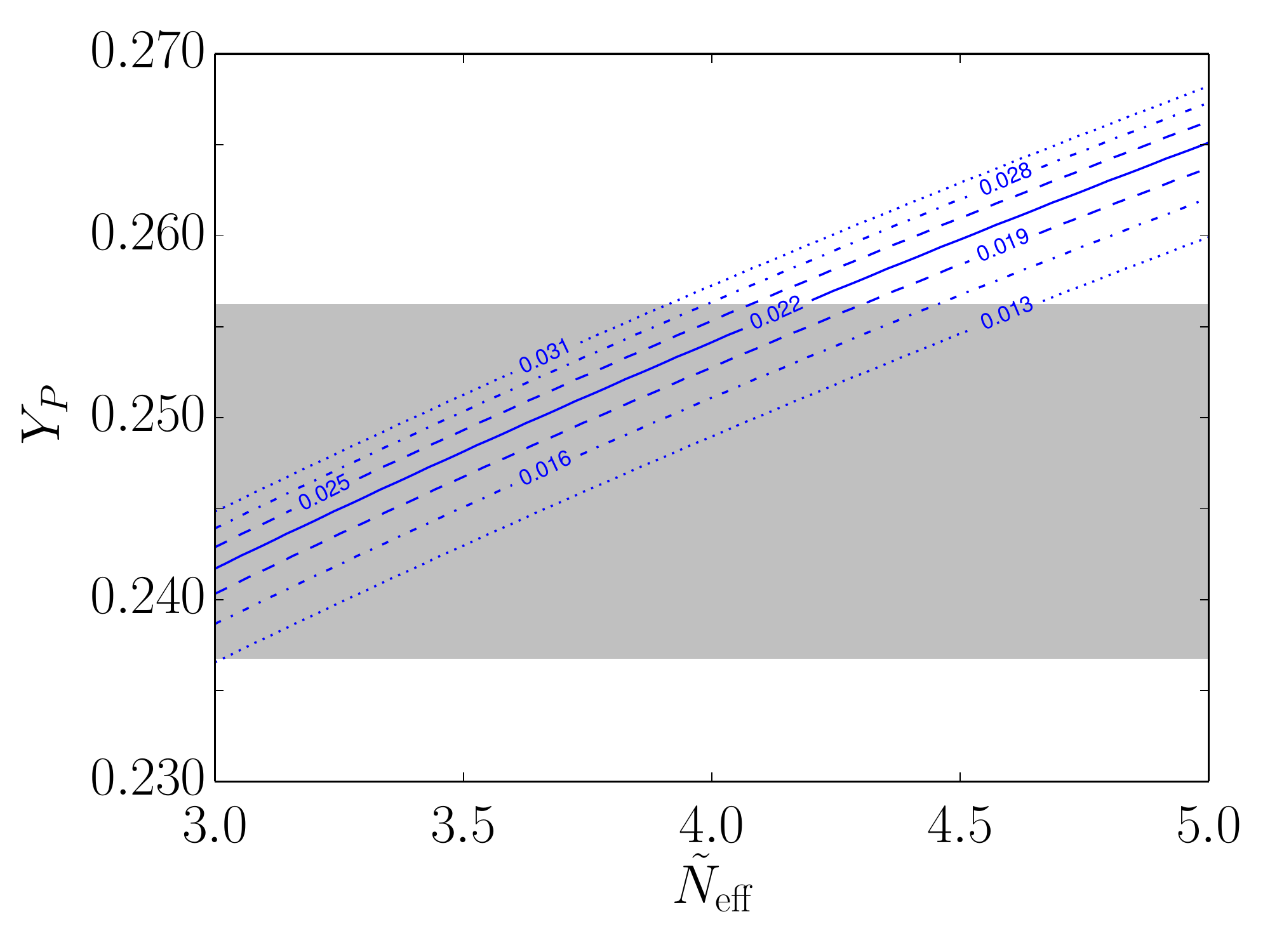}
      \includegraphics[scale=0.60]{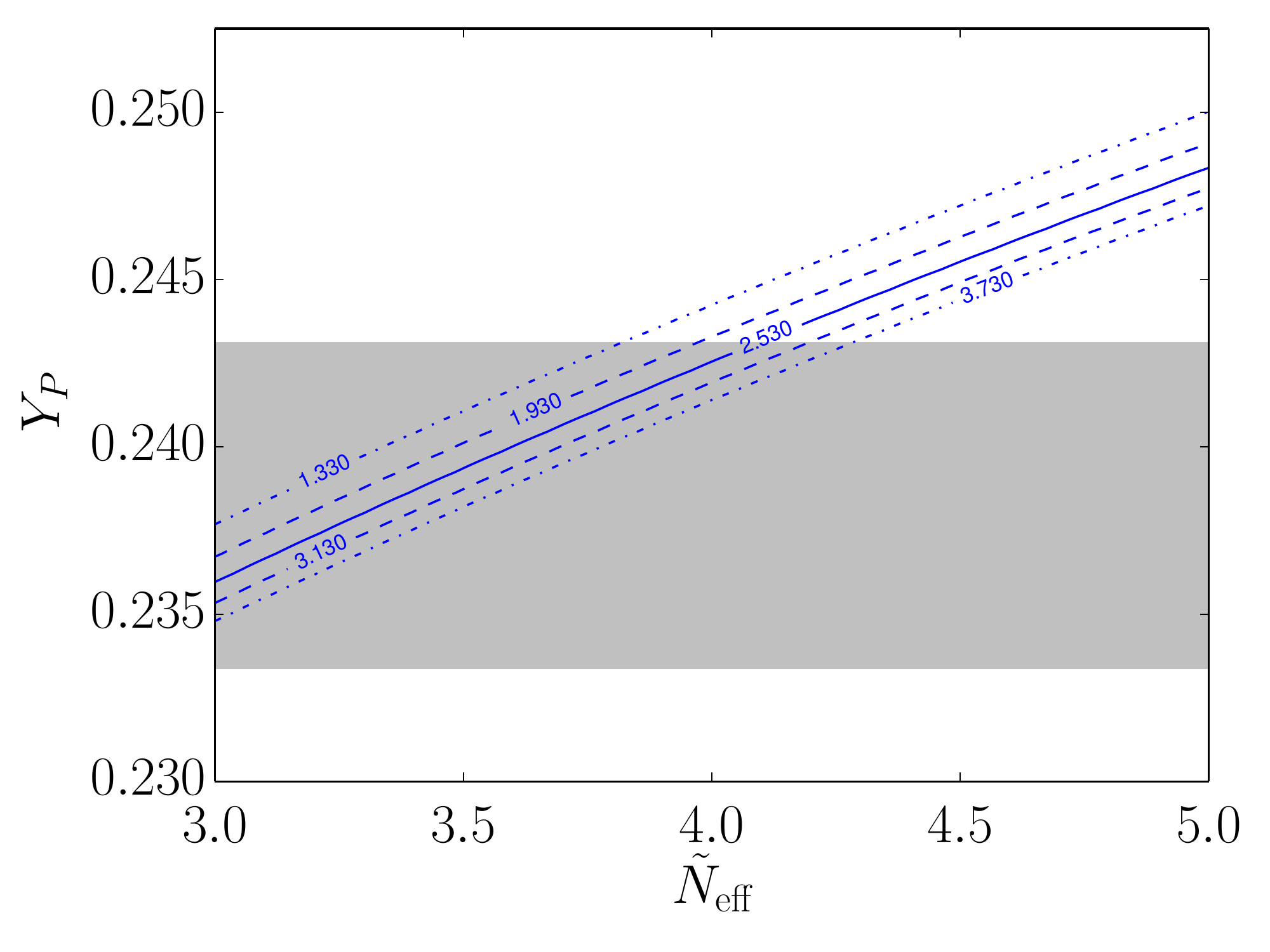}
   \end{center}
   \caption{\label{fig:Ypvn}(Top) $Y_P$
   plotted against \neffobs for contours of constant \barnum.
   The solid curve is the best-fit value of Ref.\ \cite{PlanckXVI:2013}.
   The contours are spaced by $\Delta\barnum = 0.003$.
   (Bottom) $Y_P$ versus \neffobs
   for contours of constant $10^5\times\mbox{D/H}$.
   The solid curve is the preferred value of Ref.\ \cite{Cooke:2014do}.
   The contours are spaced by $\Delta(10^5\times\mbox{D/H})=0.6$.}
\end{figure}

\begin{figure}
   \begin{center}
      \includegraphics[scale=0.60]{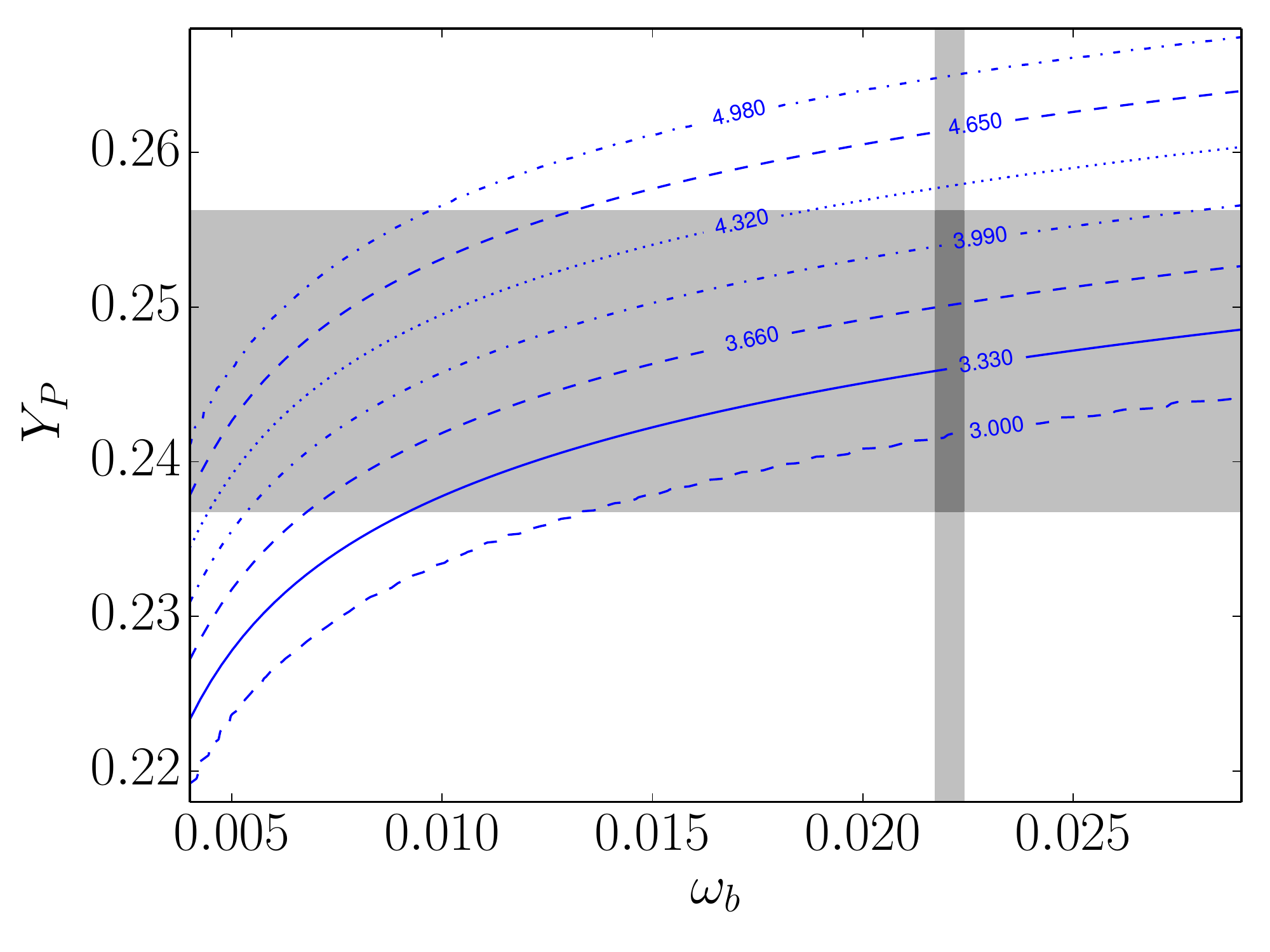}
   \end{center}
   \caption{\label{fig:Ypvbn} $Y_P$
   plotted against \barnum for contours of constant \neffobs.
   The contours are spaced by $\Delta\neffobs=0.33$.}
\end{figure}

Figures \ref{fig:nvocy}--\ref{fig:Ypvbn} show the results of
computations in which the four parameters $\omega_b$, $Y_P$, D/H, and
\neffobs are varied.
We begin by varying the two model inputs: $\omega_b$ and \neffobs.
The upper panel in Fig.\ \ref{fig:nvocy} shows the dependence of
\neffobs\ on $\omega_b$ for curves of constant $Y_P$; the vertical
band is the Planck value of $\omega_b = 0.02207 \pm 0.00033$. The
figure is generated by first choosing a value for the baryon density
in the range $0.004 \le \omega_b \le 0.029$. Each selected value of
the dark radiation parameter in the range $3 \le \neffobs \le 4.5$
allows for the prediction of $Y_P$ and D/H by parametrizing the
radiation energy density as in Eq.\ \eqref{eqn:rhor-neff}.  The values
so obtained are plotted as contours in the $\neffobs$-$\omega_b$
plane in the upper and lower panels of Fig.\ \ref{fig:nvocy}. The solid
curve is the preferred value $Y_P=0.2465 \pm 0.0097$ of
Ref.\ \cite{Aver:2013ue}, which is a selection of observations of metal
poor extragalactic H {\sc ii} regions. The contours are spaced by
roughly $0.0097/3$ showing that \neffobs\ is not strongly constrained
by values of $Y_P$ alone; this is a manifestation of the degeneracy of
\neffobs\ and $Y_P$.  For example, at $Y_P = 0.2465 \pm 0.0035$,
corresponding to the contours closest to $Y_P=0.2465$, the range
allowed \neffobs\ is nearly consistent with both the standard,
calculated value $\neff=3.046$ and the Ref.\ \cite{PlanckXVI:2013}
derived value of $\neff=3.30 \pm 0.27$.

Predictions of the primordial deuterium abundance are a much more
sensitive constraint upon allowed values of \neffobs.
This can be seen in the lower panel of Fig.\ \ref{fig:nvocy}.
The solid line
contour with $10^5\times\mbox{D/H} = 2.530 \pm 0.04$ corresponds to the
recent measurement of Ref.\ \cite{Cooke:2014do}. Contours in this figure
are separated by the one standard deviation of Ref.\ \cite{Cooke:2014do}. There are two
points of interest regarding the deuterium figure. First, as noted in
Ref.\ \cite{Cooke:2014do}, observation of the primordial component of
deuterium is precise enough to begin to constrain the microscopic
physics of the thermally averaged nuclear reaction rates and their
cross sections. Additionally, given the precision of the current and
forthcoming deuterium measurements and the strong dependence of
\neffobs\
on its value (at constant $\omega_b$), we advocate using D/H as
a prior, over $Y_P$, for future base model parameter searches.

The degeneracy between \neffobs\ and $Y_P$ is again evident in Fig.\
\ref{fig:DHvbn}.  Each plot explores the D/H vs.\ $\omega_b$ contour
space, where the upper plot contains contours of constant $Y_P$ and
the lower plot contains contours of constant \neffobs.  The shaded
bands in each figure indicate the one-sigma observations of $\omega_b$ and D/H
from Refs.\ \cite{PlanckXVI:2013,Cooke:2014do}, respectively.  Deuterium
is not an input parameter into our model.  We compute it by choosing a
baryon number and iteratively change the dark-radiation parameter,
$\delta_{\text dr}$ until matching the chosen deuterium target.  The
outputs from the process are \neffobs\ and $Y_P$.
Values of $\omega_b$, D/H and \neffobs\ are in satisfactory agreement
with the standard cosmology at the precision of current observations.

The quantities D/H and $\omega_b$ are the tightest
observationally-constrained parameters we are currently investigating.
Figure \ref{fig:Ypvn} shows two plots in the $Y_P$-$N_\text{eff}$ plane
with contours of constant $\omega_b$ (upper plot), and contours of
constant $10^5\times\mbox{D/H}$ (lower plot).
The horizontal
band in each figure indicates the one-sigma observation of $Y_P$
from Ref.\ \cite{Aver:2013ue}.
Like D/H, $Y_P$ is not
an input into our model.  Consequently, we adopt the identical
iterative method for $Y_P$ in Fig.\ \ref{fig:Ypvn} as we do for D/H in
Fig.\ \ref{fig:DHvbn}.  For the $\omega_b$ (upper) plot of
Fig.\ \ref{fig:Ypvn}, the solid contour line is the best-fit value of
Ref.\ \cite{PlanckXVI:2013} with nine-sigma spacing of the contours.
The contours exist in a subspace of the $Y_P$-$N_\text{eff}$ plane which
is well within current observations, but nevertheless could span a
range of radically different physics.  Similarly, the bottom plot
shows the $10^5\times\mbox{D/H}$ value of Ref.\ \cite{Cooke:2014do} as
the solid contour with the other contours spaced fifteen-sigma apart.
Clearly, $Y_P$ and \neffobs\ do not constrain the cosmological model
as tightly as $\omega_b$ and D/H. This observation indicates the
import of using the next generation of 30-meter class telescopes and CMB
observation to better determine the light element abundances,
particularly, $Y_P$ with high precision.

Figure \ref{fig:Ypvbn} shows how \neffobs\ changes in the
$Y_P$-$\omega_b$ plane.
The shaded
bands in each figure indicate the one-sigma observations of $\omega_b$ and $Y_P$
from Refs.\ \cite{PlanckXVI:2013,Aver:2013ue}, respectively.
We may conclude from this plot that there could
exist many different values of \neffobs\ consistent with the
observations of $Y_P$ and $\omega_b$. However, we caution against such
a conclusion without considering the effect on D/H.  In fact, we
choose not to include a figure of contours of D/H in the
$Y_P$-$\omega_b$ plane because the strong sensitivity of D/H to
$\omega_b$ produces too large a range of values for $Y_P$ to be useful
as a constraint of the cosmological model.

All calculations show a consistency between $\omega_b$, \neffobs, D/H,
and $Y_P$ to a conservative limit of two-sigma error range in each
observation.  We expect the uncertainties in each observation to
improve in the coming years with large ground-based CMB experiments
\cite{2014SPIE.9153E..1PB,2014ApJ...794..171T} and 30-meter class
telescopes \cite{TMT,GMT,EELT}.  Future high-precision measurements
may result in tensions for the best-fit values of $\omega_b$,
\neffobs, and D/H. These tensions could be indications of the need for
more precise theoretical and numerical approaches or could signal the
presence of physics beyond the standard model.  As it stands here, the
bottom panels of Figs.\ \ref{fig:nvocy} and \ref{fig:DHvbn} shows that
the tension between \barnum and D/H cannot be resolved with the
addition of extra radiation energy density.  Uncertainties in nuclear
reactions may produce disagreement between \barnum and D/H, allowing
D/H to become a probe of nuclear physics.  It is also possible that
the spectroscopic determination of D/H may be subject to small
systematic errors only recognizable at such precision.
An exciting prospect is the need to revise the CSM to resolve tensions
with the observations of D/H and \barnum, possibly leading to the
conclusion of BSM physics active during BBN.

\section{Examples for neutrino sector BSM physics}
\label{sec:bsm} 
We describe three examples of unresolved issues in BSM/CSM physics,
which require the fully self-consistent parameter determination
described in previous sections. We consider, in turn, models
incorporating neutrino rest mass, sterile neutrinos, and non-zero
lepton numbers.  We use, throughout this section, the
observationally-inferred definition of \neff, \neffobs.

\begin{figure}
   \begin{center}
      \includegraphics[scale=0.60]{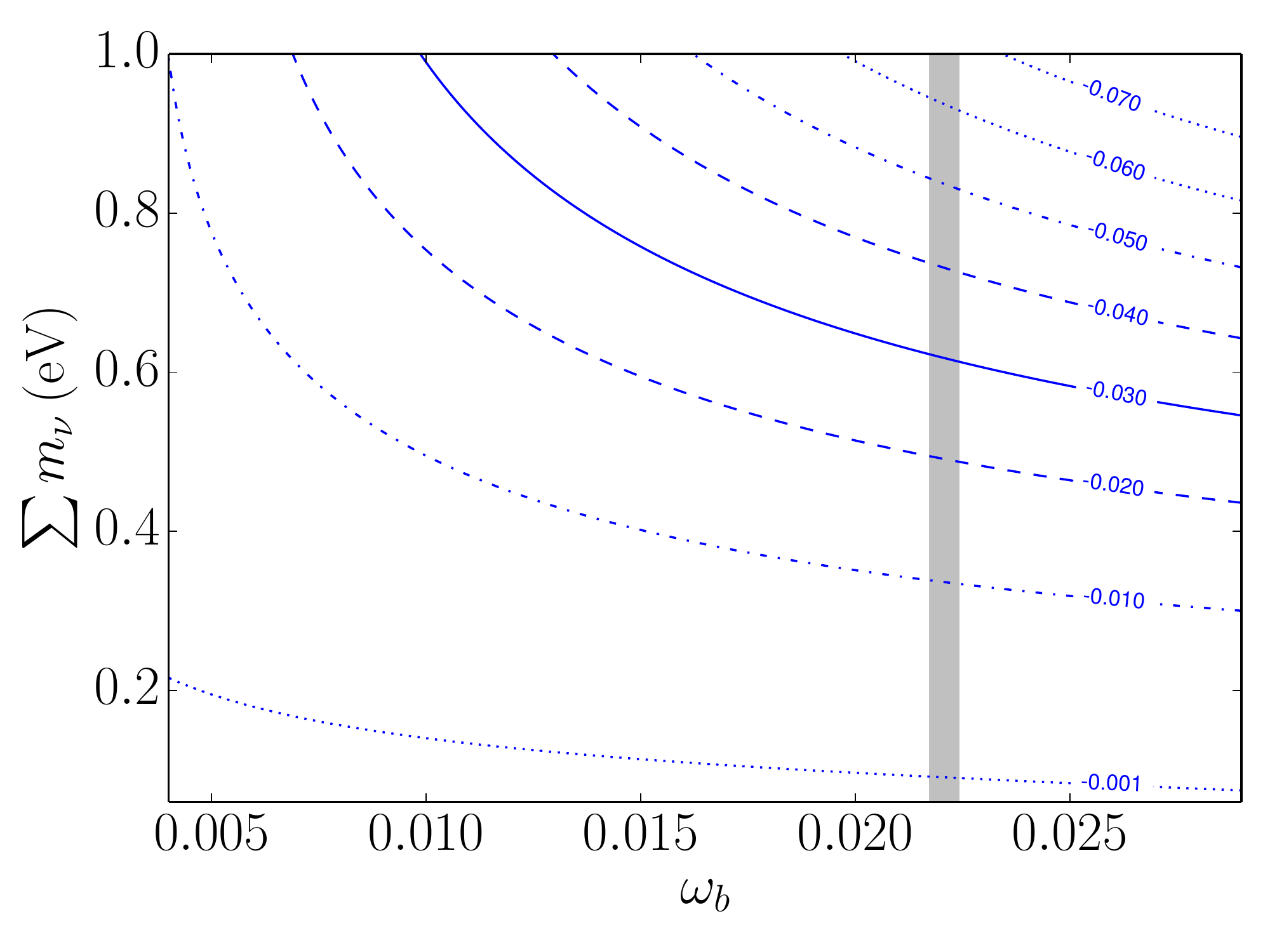}
   \end{center}
   \caption{\label{fig:neff-summnu-barnum}Determination of \summnu
   plotted against \barnum at constant $\Delta\neffobs$.
   The contours are spaced by $\approx0.01$ in values of $\Delta\neffobs$.
   All contours correspond to $\Delta\neffobs < 0$.}
\end{figure}

\subsection{Neutrino rest mass}
\label{ssec:summnu} 
Section \ref{sec:4pts} details a self-consistent treatment of the BBN
observables $Y_P$, D/H, \neffobs, and \barnum. The sum of the light
neutrino masses, denoted \summnu, has no bearing on the determination
of primordial abundances in BBN calculations due to the high
temperatures relevant there.  Here, however, we explore the epochs and
energy scales in the history of the universe associated with the
\summnu energy scale in order to investigate the relationship between
\summnu and the other four observables of interest ($\omega_b$,
\neffobs, $Y_P$ and D/H).  Specific examples of such epochs that we
might consider include the surface of last scattering ($z\sim1100$)
and the advent of LSS ($z\lesssim 10$).  We focus on the surface of
last scattering and implications for the CMB in this paper.

Reference \cite{GFP-PRL:2014mn} (hereafter GFKPI) investigates the
effect of neutrino rest mass on \neffobs using the {\sc burst} suite
of codes.  Conventional estimates based on the energy density added by
non-zero neutrino rest masses suggest an increase in \neffth.
However, using the method outlined in Sec.\ \ref{sec:method}, GFKPI
shows a decrease in \neffobs.  The decrease is due to an effect on the
recombination history stemming from an increase in the Hubble rate,
which results in a larger free-electron fraction.  This
counterintuitive result is termed the ``neutrino-mass/recombination
(\numr) effect.''
The \numr effect manifests itself only in a
self-consistent treatment, such as that employed by GFKPI.

In addition, GFKPI investigates the dependence of the \numr effect on
\barnum. We revisit this physics here in preparation for a discussion on the
effect of non-zero lepton number \lnu later, in
Sec.\ \ref{sssec:lepneff}. Increasing \barnum leads to an enhancement of
the \numr effect when \summnu is held constant, as is evident by the
curvature of the contours in Fig.\ \ref{fig:neff-summnu-barnum}.  The
enhancement is a consequence of the effect that changing \barnum has
on the the recombination history.  We might naively expect a larger
change in the Hubble rate relative to the massless neutrino case
resulting in an enhanced \numr effect for the smaller \barnum case.
This is opposite to that observed in Fig.\
\ref{fig:neff-summnu-barnum}.  This result also is counterintuitive
based on expectations from a simple scaling of the energy density and
the resulting change in the recombination
history \cite{GFP-PRL:2014mn}.

The origin of the enhancement of the \numr effect can be understood by
considering a simplification of the Boltzmann equation that determines
the recombination history [Eq.\ \eqref{eqn:xe_def}].
We take $Y_P = 0$ for the purposes of this argument since the \numr
enhancement is insensitive to $Y_P$, as we have verified numerically
for the ranges of parameters we are considering.
In this simple scenario, $X_e = X_p$ and we obtain an expression 
for the change in the free-electron fraction:
\begin{equation}
   \label{eqn:boltz_H_gs}
   \frac{dX_e}{dt} = (1 - X_e)\beta - X_e^2
   n_e^{\rm (tot)}\alpha^{(2)},
\end{equation}
where $\beta\equiv\alpha^{(2)}(m_eT/2\pi)^{3/2}e^{-\Delta Q/T}$ is the
ionization coefficient and $\alpha^{(2)}$ is the recombination
coefficient with $\Delta Q=13.6$ eV.

Equation \eqref{eqn:boltz_H_gs} for the
recombination history shows that the free-electron disappearance rate
is proportional to the total electron number density, which in
turn is proportional to \barnum through Eq.\ \eqref{eqn:netot_def}. Equation
\eqref{eqn:netot_def} also shows how $n_e^{\rm (tot)}$ relates to
$Y_P$. Note that \barnum and $Y_P$ affect $n_e^{\rm (tot)}$
differently. However, due to the relative insensitivity of $Y_P$ to
\barnum, the \barnum dependence dominates in Eq.\ \eqref{eqn:netot_def}.
The increase in energy density from $\summnu\ne0$ and the increase in
the free-electron disappearance rate combine to alter the
recombination history so as to enhance the \numr effect for increasing
\barnum.

\subsection{Sterile neutrinos}\label{sec:stnu}
We next consider the possibility that there exists either single or
multiple sterile-neutrino species, which could have profound
implications in cosmology. We entertain two possibilities of either
light or heavy sterile neutrinos.

\subsubsection{Light sterile neutrinos}\label{ssec:lstnu} 
Observations of neutrino events in large scintillating detectors may
have revealed anomalies that could be interpreted as sterile neutrinos
with rest masses $m_{\nu_s}\sim1$ eV \cite{LSND:1997, MiniBooNE:2010, DoubleChooz:2012}.
We investigate the presence of a single
sterile neutrino in the early universe by employing a model where the
sterile state populates a thermal Fermi-Dirac shaped distribution with
temperature parameter $T_s$, possibly through flavor mixing.  The
sterile neutrino temperature, $T_s$ is taken to be less than or equal
to the active neutrino temperature $T_\nu$. The ratio $T_s/T_\nu$ is
assumed to be the same throughout weak decoupling, BBN, and
recombination.  For this analysis, we do not investigate smaller active-sterile
neutrino mixing angles with resultant non Fermi-Dirac-shaped energy
spectra \cite{KF:2008PRD, KF:2008PRL}.
Future work will consider such physics \cite{GFG:2015}.

\begin{figure}
   \begin{center}
      \includegraphics[scale=0.60]{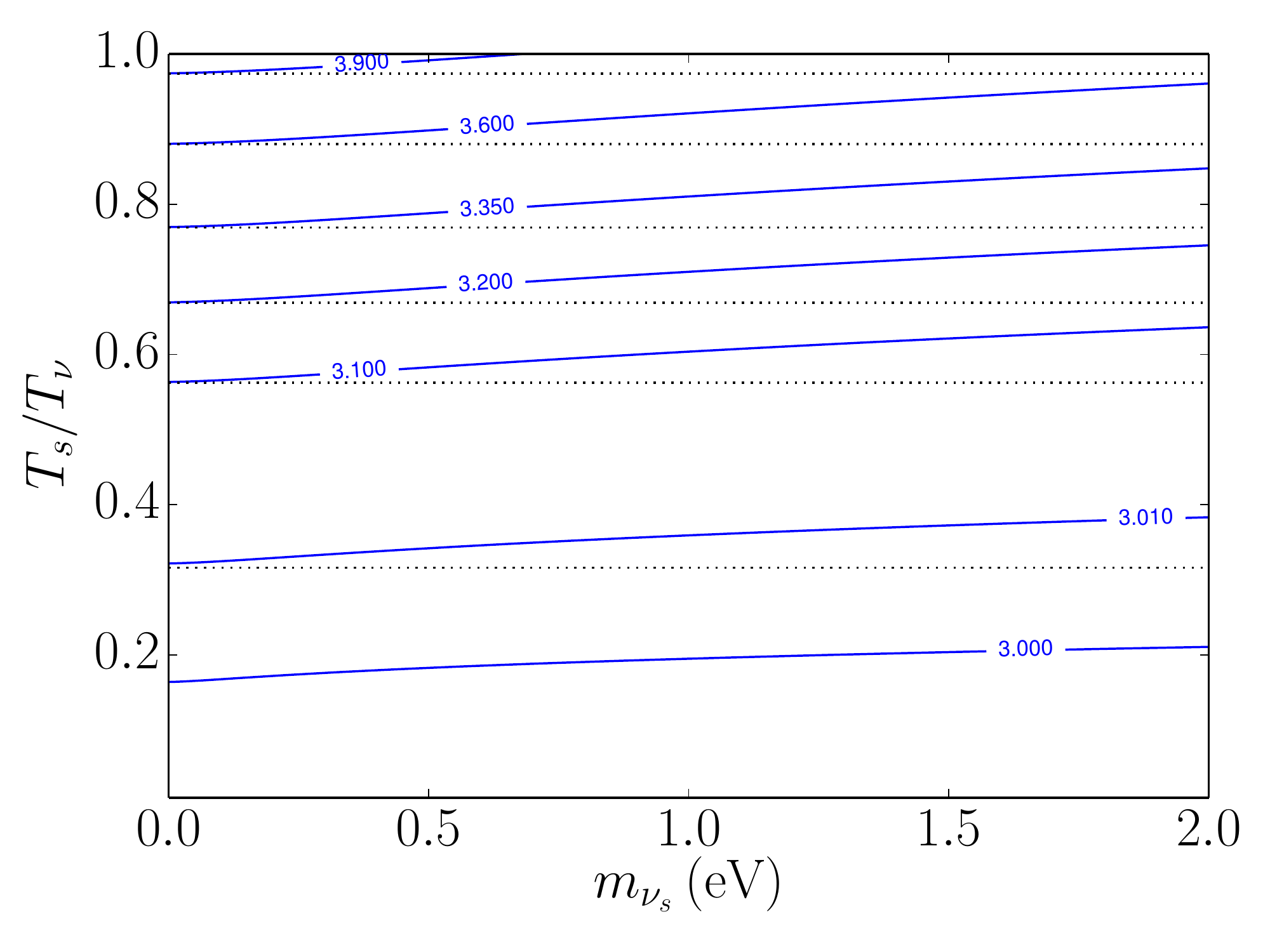}
   \end{center}
   \caption{\label{fig:stnu} Ratio of the sterile to active neutrino
      temperatures, $T_s/T_\nu$, plotted against $m_{\nu_s}$ for
      contours of constant \neffobs for $\summnu=0.06$ eV.  Horizontal
      dotted lines show the prediction if the sterile neutrino was
   massless, i.e.\ $m_{\nu_s} = 0$}
\end{figure}

Figure \ref{fig:stnu} displays contours of constant \neffobs.  The
vertical axis is the ratio $T_s/T_\nu$ and the horizontal axis the
sterile neutrino rest mass $m_{\nu_s}$.  We maintain the ratio
$T_\nu/T=(4/11)^{1/3}$, assuming covariant conservation of
entropy, starting at the end of the epoch of $e^\pm$-annihilation and
continuing throughout the remainder of the history of the universe.
The dotted lines show the expectation from the dark radiation analysis
of Sec.  \ref{sec:4pts} without employing the self-consistent,
iterative approach developed in Sec.\ \ref{sec:method}.  The deviation
of the contours from the dotted lines is again due to an effect
similar to the \numr effect but, in this instance, due to the sterile
state.
Fig.\ \ref{fig:stnu} takes the sum of the active neutrino masses to be 0.06 eV.
This is
inconsequential for large $T_s/T_\nu\lesssim1$.  For $T_s/T_\nu\lesssim0.1$,
$\Delta\neffobs < 0$ due to the \numr effect in the active neutrino
sector. As a consequence, the contour for $\neffobs=3$ is not coincident with the
$m_{\nu_s}$ axis.  Since $m_{\nu_s}$ is too small to be of any
significant kinematic effect during BBN, we need only compute BBN once for
a given value of $T_s/T_\nu$.  During recombination, $m_{\nu_s}$ is
kinematically important and affects the Hubble rate.  Hence, for every
point in the $T_s/T_\nu$-$m_{\nu_s}$ plane of Fig.\ \ref{fig:stnu} we
calculate recombination.  This figure clearly emphasizes the need for a
self-consistent treatment between BBN and recombination when
considering this BSM physics. 

\subsubsection{Heavy sterile neutrinos}
\label{ssec:hstnu} 

Heavy sterile neutrinos that decay out of equilibrium in the early
universe can affect weak decoupling and, as a consequence, primordial
nucleosynthesis \cite{2002NuPhB.632..363D, FKK:2011di,
DarkRadScherrer:2012}.  Sterile neutrinos in the rest mass range $0.1 \mbox{
GeV} \le m_{\nu_s} \le 1.0$ GeV, with lifetimes $\gtrsim1\mbox{ s}$
decaying during the weak decoupling, weak freeze-out, and/or BBN
epochs can have constrainable, sometimes dramatic, cosmological
effects.

Such sterile neutrinos have mass and vacuum mixings with $\nu_e,
\nu_\mu, \nu_\tau$ constrained by accelerator and other laboratory
oscillation experiments/observations \cite{2008RPPh...71h6201O,
   2008JPhCS.136b2031D, 2005JHEP...11..028K, 1983PhLB..128..361B,
1986PhLB..166..479B, 1988PhLB..203..332B, 1993PhLB..302..336B,
2001NuPhS..98...37N}, beta-decay experiments
\cite{2014arXiv1410.7684M}, and cosmological considerations, including
constraints on \summnu and \neff\ \cite{2003PhRvL..91x1301K,
2004JCAP...04..008H, 2009PhRvD..80l3509D, 2010JCAP...08..001H,
2010PhRvD..82h7302A, 2011PhRvD..83k5023G}.  In fact, stringent
constraints can be obtained from \neff\ limits alone
\cite{FKK:2011di}, as sterile neutrinos decaying out of equilibrium can
lead to dilution (entropy production) which, in the weak
decoupling epoch, can lead to distortions in the relic neutrino energy
spectrum, affecting \summnu, and have significant impact on the
relativistic energy content and, hence, \neff.
A sophisticated theoretical and computational treatment of
dilution physics is a challenging endeavor.
Even though these heavy
sterile neutrinos may decay away before an epoch where $T\sim10\mbox{
keV}$, they can nevertheless alter the relationship between $Y_P$,
D/H, \neff, \summnu, and \barnum, necessitating the need for a
self-consistent treatment between the weak decoupling, weak freeze-out,
BBN, recombination, photon decoupling, and advent of LSS epochs \cite{GFKPV-nuxport}.

\subsection{Lepton numbers}
\label{sec:lep} 
We examine how lepton numbers affect the primordial abundances
and \neffobs.  We define the lepton number $L_\nu$ for a neutrino
species $\nu$ in a flavor eigenstate as
\begin{equation}
\label{eqn:lnu}
L_\nu\equiv\frac{n_{\nu} - n_{\bar{\nu}}}{n_\gamma},
\end{equation}
where $n_\nu$ is the number density of neutrino species $\nu$,
$n_{\bar{\nu}}$ is the number density of anti-neutrino species
$\bar{\nu}$, and $n_\gamma$ is the number density of photons.  Here,
for illustrative purposes, we take
$L_{\nu_e}=L_{\nu_\mu}=L_{\nu_\tau}\equiv L_{\nu}$.  In fact, flavor
oscillations will bring the various lepton numbers into near equality
\cite{1991ApJ...368....1S, 2002PhRvD..66a3008A, 2002PhRvD..66b5015W}.

We use the comoving-invariant neutrino degeneracy parameter
$\xi_\nu\equiv\mu_\nu/T_\nu$, where $\mu_\nu$ is the chemical
potential of neutrino $\nu$, to compute \lnu.
The present model assumes
that the lepton number evolves
through the epoch of $e^\pm$ annihilation only in response to the relative increase
in $n_\gamma$. We do not consider any BSM physics which could alter
the difference $n_\nu - n_{\bar{\nu}}$; that is, we fix
$\xi_\nu$ throughout weak decoupling, BBN, and photon decoupling.  We
relate the degeneracy parameter to the lepton number using the
following expression \cite{KF:2008PRD}:
\begin{equation}\label{eqn:lepdeg}
  \lnu = \frac{4}{11}\frac{1}{12\zeta(3)}\left(\pi^2\xi_\nu +
  \xi_\nu^3\right),
\end{equation}
where $\zeta(3)\approx1.202$.
The factor $4/11$ in Eq.\eqref{eqn:lepdeg} implies that our
lepton numbers refer to the post $e^\pm$ annihilation epoch, where
$T_\nu/T=(4/11)^{1/3}$.

\begin{figure}
   \begin{center}
      \includegraphics[scale=0.60]{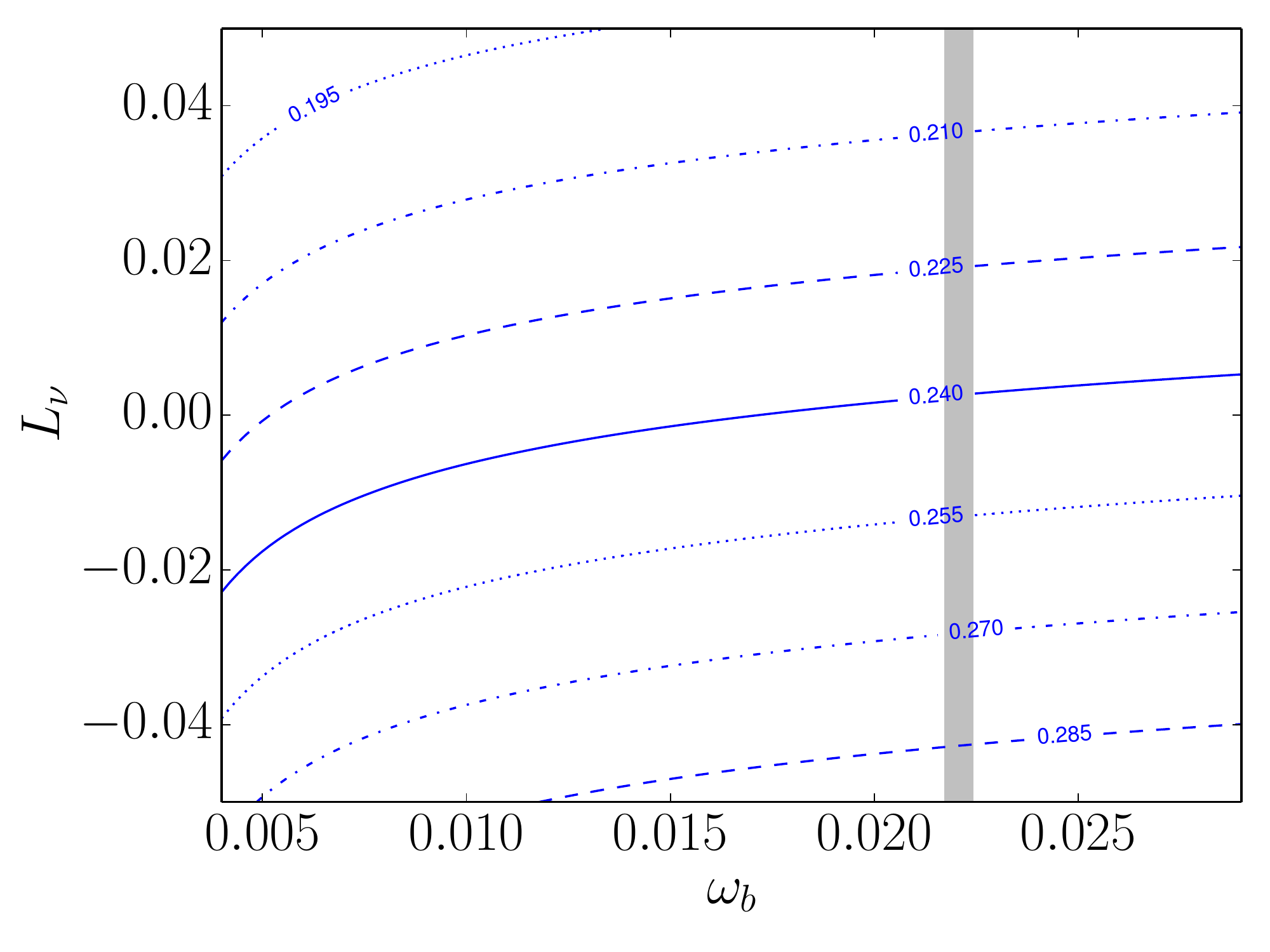}
   \end{center}
   \caption{\label{fig:ypxivbn}Lepton asymmetry $L_\nu$
   [Eq.\eqref{eqn:lnu}] plotted against \barnum for contours of
constant $Y_P$.  The contours are spaced by $\Delta Y_P = 0.015$.}
\end{figure}

\begin{figure}
   \begin{center}
      \includegraphics[scale=0.60]{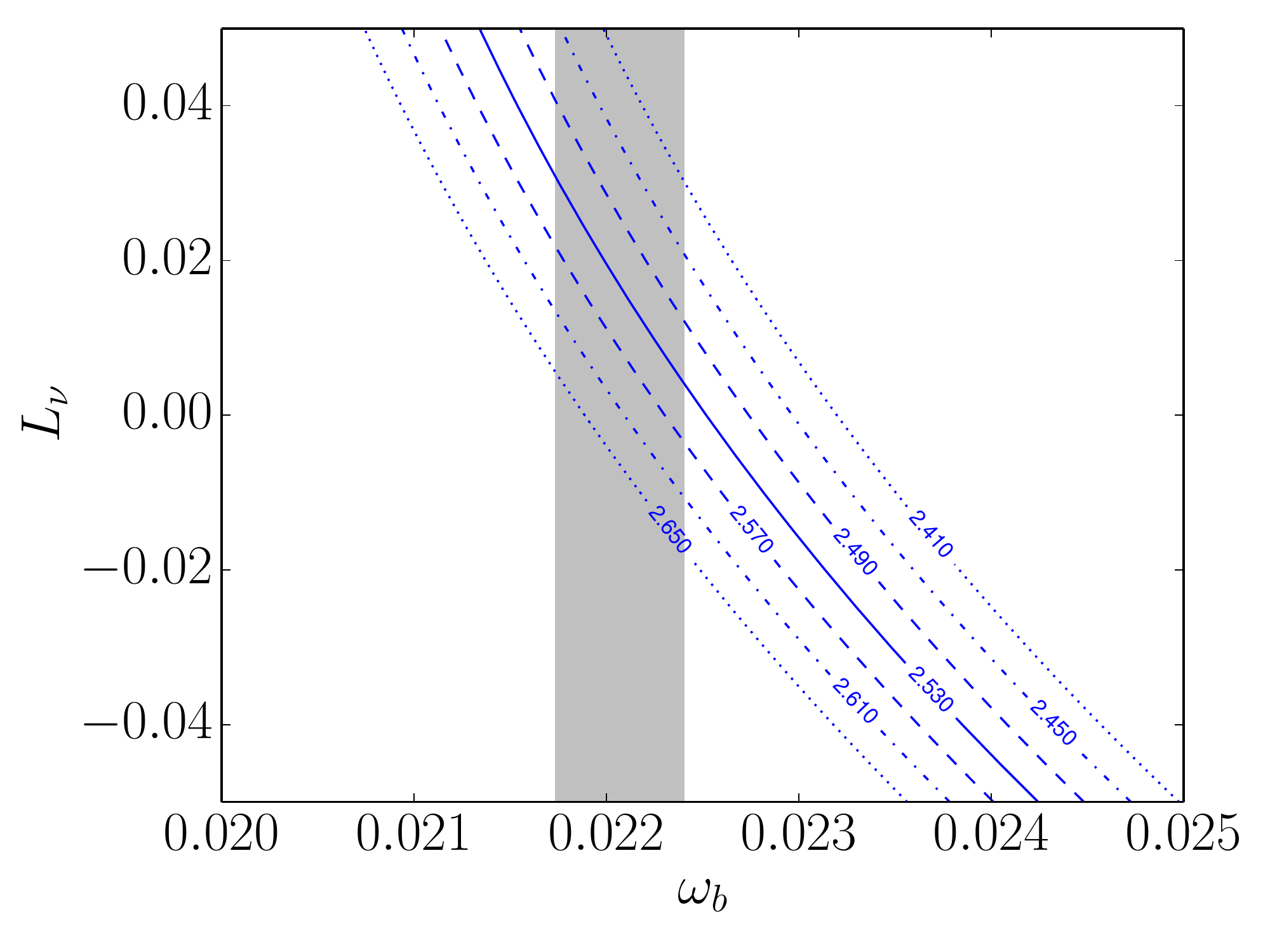}
   \end{center}
   \caption{\label{fig:dhxivbn}Lepton asymmetry $L_\nu$ plotted
   against \barnum for contours of constant $10^5\times{\rm D/H}$.
   The solid curve is the preferred value of Ref.\ \cite{Cooke:2014do}.
   The contours are spaced by $\Delta(10^5\times\mbox{D/H})=0.04$.}
\end{figure}

\subsubsection{Effect on nucleosynthesis}
The helium mass fraction is sensitive to the neutron-to-proton ratio,
$n/p$.  We determine $n/p$ by calculating the weak rates associated
with neutrino-nucleon reactions, namely:
\begin{align}
  \label{eqn:urca1}\nu_e + n &\rightleftharpoons e^- + p^+\\
  \label{eqn:urca2}\bar{\nu}_e + p^+ &\rightleftharpoons e^+ + n
\end{align}
and in addition, neutron and inverse neutron decay:
\begin{equation}\label{eqn:ndecay}
  n \rightleftharpoons e^- + \bar{\nu}_e + p^+
\end{equation}
The net rates in reactions \eqref{eqn:urca1} and \eqref{eqn:urca2} fall below
the Hubble rate at the epoch of weak freeze-out.
Weak freeze-out largely precedes the alpha-particle formation process
in BBN, though unlike the brief time/temperature range of
$\alpha$-formation, weak freeze-out occurs over several Hubble times
at this epoch. The rates in reactions \eqref{eqn:urca1} through
\eqref{eqn:ndecay} are sensitive to the neutrino and $e^\pm$
distributions. We follow Ref.\ \cite{SMK:1993bb} to evolve $T$
and the electron chemical potential in order to maintain equilibrium
between the electrons, positrons and photons.  For the electron-flavor
neutrinos, we use the comoving invariants $aT_\nu$ and $\xi_{\nu_e}$
to compute the neutrino distributions.  We set $\summnu=0$ as
neutrinos of sub-eV rest mass remain ultra-relativistic throughout
weak freeze-out.

Figures \ref{fig:ypxivbn} and \ref{fig:dhxivbn} show the helium mass
fraction and the relative deuterium abundance, respectively.  Each
plot is in the \lnu-\barnum plane for contours of constant primordial
abundance.  The relationships between lepton number and nucleosynthesis
are well known \cite{2004NJPh....6..117K,2006PhRvD..74h5008S}.
Increasing $L_{\nu_e}$ leads to an overabundance of neutrinos compared
to anti-neutrinos.  The forward rate of reaction \eqref{eqn:urca1}
freezes-out after the forward rate of reaction \eqref{eqn:urca2}.  The
imbalance lowers $n/p$ which lowers $Y_P$ as seen in Fig.\
\ref{fig:ypxivbn}.  The decrease in $n/p$ also leads to a decrease in
D/H, although deuterium is not as sensitive to \lnu as helium.
However, D/H is known to much higher precision than is $Y_P$.

Comparing with recent observations \cite{Aver:2013ue, Cooke:2014do},
the two light element abundances achieve consistency at $2\sigma$.
$Y_P$ prefers a value of $\lnu<0$ whereas D/H prefers a positive value
of \lnu.  If future observations of the light-element abundances were
to show a larger disagreement than $2\sigma$, lepton numbers of
identical value could not solely rectify the tension.  Future analyses
will consider scenarios with multiple facets of BSM physics including
non-zero lepton numbers \cite{AFG:2015}.  This analysis will use
\neffobs as a discriminating factor.

\subsubsection{Effect on \neff}
\label{sssec:lepneff} 
\begin{figure}
   \begin{center}
   \includegraphics[scale=0.60]{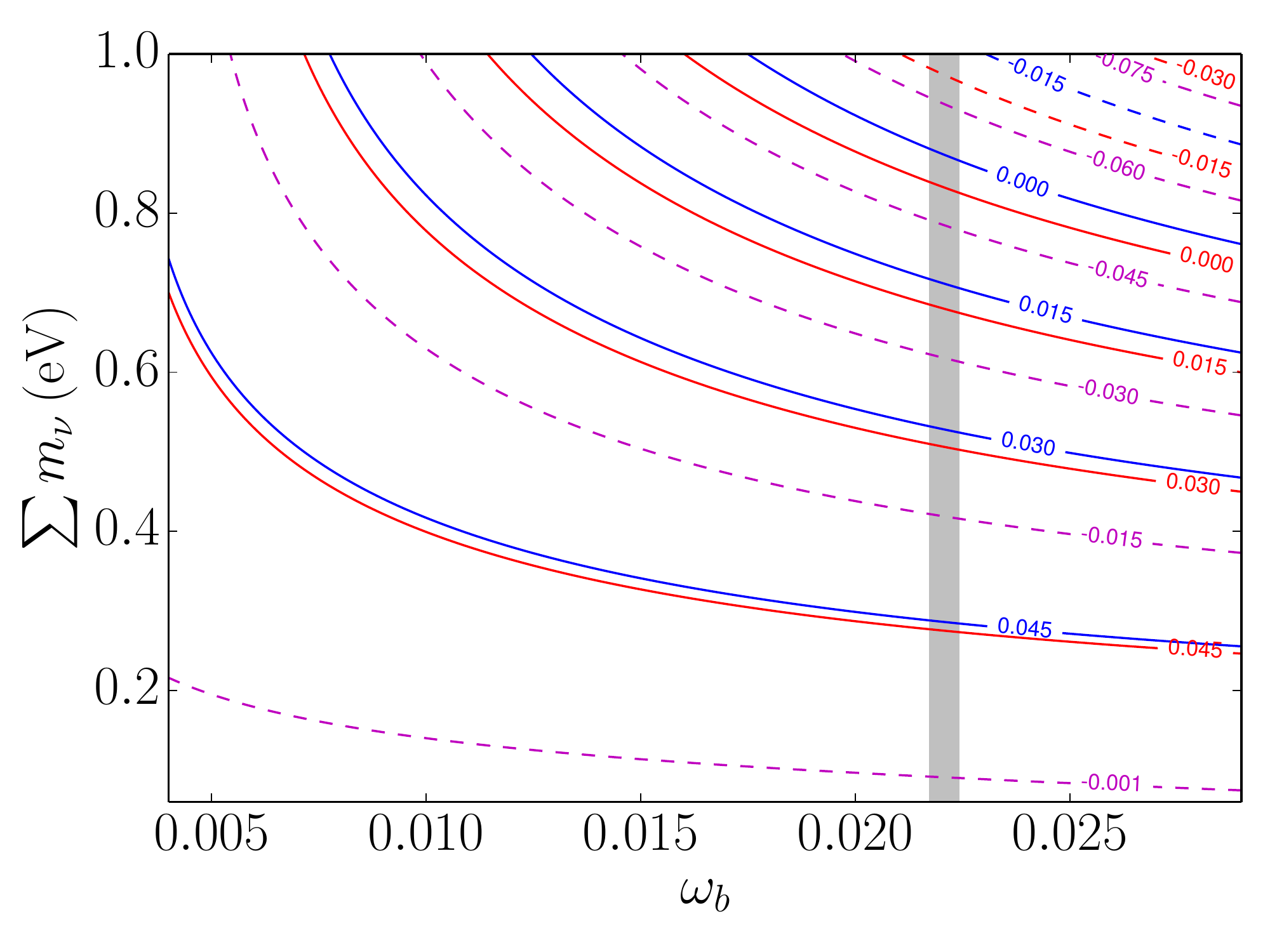}
   \end{center}
   \caption{\label{fig:3lep}\summnu plotted against \barnum for
      contours of constant $\Delta\neffobs$. The blue contours are for
      $\lnu=-0.05$.  The magenta contours are for $\lnu=0$.  The red
      contours are for $\lnu=0.05$.  Solid contours are for positive
      values; dashed contours are for negative values.}
\end{figure}

We consider how two aspects of non-CSM/BSM physics ($\lnu\ne 0$ and/or
$\summnu\ne0$) modify \neffobs.
If we set the neutrino
rest mass to $\summnu\ne0$, we can investigate whether the \numr
effect still applies with non-zero \lnu. Figure \ref{fig:3lep}
shows the changes to \neffobs in the \summnu-\barnum plane for three
values of $\lnu=-0.05,0,0.05$ corresponding to blue, magenta, and red
contours, respectively.  The non-zero lepton number increases
\neffobs for small values of \summnu in accordance with
Ref.\cite{Shimon:2010cb}.  However, for values of $\summnu\sim1.0$ eV,
the \numr effect overwhelms the extra energy density from more
particles to lower \neffobs below three.

Figure \ref{fig:3lep} also shows that, despite the
total energy density being insensitive to the sign of $L_\nu$,
the contours for non-zero values of \lnu with opposite sign
do not overlap because $Y_P$ depends sensitively on its value.
$Y_P$ is largest for the blue contours, so more helium
suppresses the \numr effect.

Note that taking
$\lnu\ne0$ conflates the interpretation that the effect of \summnu
is identical to perturbations in the matter power spectrum, which
are used in calculating the suppression of power on small scales.
This is borne out by the present model where the \summnu statistic
cannot be equated to the cosmological measurement. In our model,
\summnu is simply the sum of the active vacuum neutrino mass
eigenvalues. The observationally determined value of \summnu
depends on quantities other than the sum of the active neutrino
masses such as their energy distributions.

\section{Conclusions and outlook}
\label{sec:outl} 

Cosmological considerations are a key route to exploring BSM physics. This is
especially true for the neutrino sector, where there are many outstanding questions
and where laboratory experiments are limited in what aspects of this physics can be
addressed. In this paper we have argued that a self consistent treatment of BSM
issues, across all epochs from weak decoupling to photon decoupling, is the best way
to take advantage of the expected coming increase in precision of CMB measurements
and observationally-inferred primordial abundances of the light elements.
We employ a limited prescription to link the salient features of self consistency
between early-time neutrino dynamics and the surface of last scattering.
We couple the weak decoupling
and nucleosynthesis of early times to CMB observables, including
baryon-to-photon ratio (equivalently \barnum), sound horizon, and photon diffusion length.

We have shown that such a self consistent treatment is necessary, in part because
new neutrino physics can alter the relationships between different cosmological epochs.
For example, \barnum and
other CMB observables affect the calculated yields of deuterium and helium.
In addition, the
calculated relic neutrino energy spectra after weak decoupling affects the predicted value
of \neffobs at photon decoupling.
The principal tool in our analysis is the suite of {\sc burst} codes for
nucleosynthesis and neutrino interactions and energy transport.

A case in point is our investigation of the relationship between neutrino rest
masses, i.e.\ \summnu, and the four potential observables \barnum,
\neffobs, $Y_P$, and D/H. This analysis reveals the
``neutrino-mass/recombination'' (\numr) effect first described in
Ref.\ \cite{GFP-PRL:2014mn}. The \numr effect is below the threshold of current CMB
capabilities, but may not be in future observations \cite{Abazajian201566}.

There have been spectacular advances in the measurements/observations
of neutrino properties. We know the neutrino mass-squared differences
and three of the four parameters in the unitary transformation between
the energy eigenstates (mass states) and the weak interaction
eigenstates (flavor states) of the three active neutrinos (only the
$CP$-violating phase remains unmeasured).
As active neutrinos mix in vacuum and have non-zero rest masses, the
question indubitably arises of whether there exist ``sterile''
neutrino states.
If indeed sterile neutrinos do exist, we acknowledge that the parameter
space of mass, vacuum mixing angle, and number is enormous.  However,
sterile neutrinos could have profound effects in all of the epochs
under study in this paper. This possibility makes a self consistent
treatment of these effects a powerful basis for constraining sterile
neutrino states.

In this paper we have considered scenarios for both ``light'' (mass $\sim
1\,{\rm eV}$) and ``heavy'' (mass $\sim 0.1 - 1\,{\rm GeV}$) sterile
neutrinos.  In the former case we consider cases where the sterile neutrino relic
energy spectra are Fermi-Dirac black-body shaped, though with a
temperature parameter $T_s$ differing from that characterizing the relic
energy spectra of active neutrino species. We show here that the \numr effect has
interesting consequences and that this case {\it demands} a self consistent
treatment of recombination and BBN. Additionally, heavy sterile neutrino decay out of equilibrium
can lead to dilution and high energy relic active neutrinos, and both of these
features potentially can have dramatic and constrainable effects on CMB-epoch
observables.
This implies that CMB observations can indirectly probe the C$\nu$B
and explore active-sterile
mass/vacuum-mixing parameter space unavailable to current accelerator-based
experiments.

We have also studied the effects of non-zero lepton numbers on the relationship
between CMB observables, nucleosynthesis, and neutrino physics. Our conclusion is
that the primordial deuterium abundance is a potentially powerful probe of lepton
number. However, an eventual CMB-{\it only}
measurement of the primordial helium abundance $Y_P$ will be the most powerful
probe of lepton number and many other issues. Determining $Y_P$ from CMB
observables will require a sophisticated self-consistent approach to BBN, neutrino
physics, and photon decoupling transport physics.

Finally, our study has revealed a potential tension between \neffobs,
\barnum, and the primordial deuterium abundance, D/H, inferred from
high redshift QSO absorption systems. In fact, if the advent of 30-m class
telescopes in the near future allows for a decrease in errors in
observationally-inferred D/H to the $\sim 1\,\%$ level, while observed \barnum
and \neffobs maintain their respective current central values, then
tension is unavoidable. This may signal BSM or non-CSM physics, likely in the
neutrino sector, or it could point to not understanding systematics in the damped
Lyman-$\alpha$ cloud measurements of the isotope-shifted hydrogen absorption lines.
We advocate using future instruments to explore the rich physics of weak decoupling,
nucleosynthesis, and photon decoupling to discover
what role BSM neutrino physics has in these epochs.

\clearpage 
\acknowledgments 
We would like to acknowledge the Institutional Computing Program at
Los Alamos National Laboratory for use of their HPC cluster resources.
EG acknowledges the San Diego Supercomputer Center for their use
of HPC resources and helpful technical support.
This work was supported in part by NSF grant PHY-1307372 at UC San
Diego, by the Los Alamos National Laboratory Institute for Geophysics,
Space Sciences and Signatures subcontracts, and the National Nuclear
Security Administration of the U.S.\ Department of Energy at Los
Alamos National Laboratory under Contract No.\ DE-AC52-06NA25396.  We
thank Lauren Gilbert, Jeremy Ariche, and J.J.\ Cherry for helpful
discussions.

\bibliographystyle{JHEP}
\bibliography{master}

\providecommand{\href}[2]{#2}\begingroup\raggedright\begin{thebibliography}{10}

\bibitem{WMAP:2013ny}
G.~{Hinshaw} et~al., {\it {Nine-year Wilkinson Microwave Anisotropy Probe
  (WMAP) Observations: Cosmological Parameter Results}},  {\em \apjs} {\bf 208}
  (Oct., 2013) 19.

\bibitem{ACT:2013cp}
J.~L. {Sievers} et~al., {\it {The Atacama Cosmology Telescope: cosmological
  parameters from three seasons of data}},  {\em \jcap} {\bf 10} (Oct., 2013)
  60.

\bibitem{SPT:2011hl}
R.~{Keisler} et~al., {\it {A Measurement of the Damping Tail of the Cosmic
  Microwave Background Power Spectrum with the South Pole Telescope}},  {\em
  \apj} {\bf 743} (Dec., 2011) 28.

\bibitem{PlanckI:2013ov}
{Planck Collaboration}, P.~A.~R. {Ade}, N.~{Aghanim}, M.~I.~R. {Alves},
  C.~{Armitage-Caplan}, M.~{Arnaud}, M.~{Ashdown}, F.~{Atrio-Barandela},
  J.~{Aumont}, H.~{Aussel}, and et~al., {\it {Planck 2013 results. I. Overview
  of products and scientific results}},  {\em ArXiv e-prints} (Mar., 2013)
  [\href{http://xxx.lanl.gov/abs/1303.5062}{{\tt arXiv:1303.5062}}].

\bibitem{Izotov:2010ns}
Y.~I. {Izotov} and T.~X. {Thuan}, {\it {The Primordial Abundance of $^{4}$He:
  Evidence for Non-Standard Big Bang Nucleosynthesis}},  {\em \apjl} {\bf 710}
  (Feb., 2010) L67--L71.

\bibitem{Pettini:2012yd}
M.~{Pettini} and R.~{Cooke}, {\it {A new, precise measurement of the primordial
  abundance of deuterium}},  {\em \mnras} {\bf 425} (Oct., 2012) 2477--2486.

\bibitem{Aver:2013ue}
E.~{Aver}, K.~A. {Olive}, R.~L. {Porter}, and E.~D. {Skillman}, {\it {The
  primordial helium abundance from updated emissivities}},  {\em \jcap} {\bf
  11} (Nov., 2013) 17.

\bibitem{Cooke:2014do}
R.~J. Cooke, M.~Pettini, R.~A. Jorgenson, M.~T. Murphy, and C.~C. Steidel, {\it
  Precision measures of the primordial abundance of deuterium},  {\em The
  Astrophysical Journal} {\bf 781} (2014), no.~1 31.

\bibitem{Shimon:2010cb}
M.~{Shimon}, N.~J. {Miller}, C.~T. {Kishimoto}, C.~J. {Smith}, G.~M. {Fuller},
  and B.~G. {Keating}, {\it {Using Big Bang Nucleosynthesis to extend CMB
  probes of neutrino physics}},  {\em \jcap} {\bf 5} (May, 2010) 37.

\bibitem{Hamann:2008bc}
J.~{Hamann}, J.~{Lesgourgues}, and G.~{Mangano}, {\it {Using big bang
  nucleosynthesis in cosmological parameter extraction from the cosmic
  microwave background: a forecast for PLANCK}},  {\em \jcap} {\bf 3} (Mar.,
  2008) 4.

\bibitem{PlanckXVI:2013}
{Planck Collaboration}, P.~A.~R. {Ade}, N.~{Aghanim}, C.~{Armitage-Caplan},
  M.~{Arnaud}, M.~{Ashdown}, F.~{Atrio-Barandela}, J.~{Aumont},
  C.~{Baccigalupi}, A.~J. {Banday}, and et~al., {\it {Planck 2013 results. XVI.
  Cosmological parameters}},  {\em ArXiv e-prints} (Mar., 2013)
  [\href{http://xxx.lanl.gov/abs/1303.5076}{{\tt arXiv:1303.5076}}].

\bibitem{VFC:QKE}
A.~{Vlasenko}, G.~M. {Fuller}, and V.~{Cirigliano}, {\it {Neutrino quantum
  kinetics}},  {\em \prd} {\bf 89} (May, 2014) 105004,
  [\href{http://xxx.lanl.gov/abs/1309.2628}{{\tt arXiv:1309.2628}}].

\bibitem{2013PhRvD..87k3010V}
C.~{Volpe}, D.~{V{\"a}{\"a}n{\"a}nen}, and C.~{Espinoza}, {\it {Extended
  evolution equations for neutrino propagation in astrophysical and
  cosmological environments}},  {\em \prd} {\bf 87} (June, 2013) 113010,
  [\href{http://xxx.lanl.gov/abs/1302.2374}{{\tt arXiv:1302.2374}}].

\bibitem{2014PhRvD..90l5040S}
J.~{Serreau} and C.~{Volpe}, {\it {Neutrino-antineutrino correlations in dense
  anisotropic media}},  {\em \prd} {\bf 90} (Dec., 2014) 125040,
  [\href{http://xxx.lanl.gov/abs/1409.3591}{{\tt arXiv:1409.3591}}].

\bibitem{Lewis:1999bs}
A.~Lewis, A.~Challinor, and A.~Lasenby, {\it Efficient computation of {CMB}
  anisotropies in closed {FRW} models},  {\em Astrophys. J.} {\bf 538} (2000)
  473--476, [\href{http://xxx.lanl.gov/abs/astro-ph/9911177}{{\tt
  astro-ph/9911177}}].

\bibitem{FKK:2011di}
G.~M. {Fuller}, C.~T. {Kishimoto}, and A.~{Kusenko}, {\it {Heavy sterile
  neutrinos, entropy and relativistic energy production, and the relic neutrino
  background}},  {\em ArXiv e-prints} (Oct., 2011)
  [\href{http://xxx.lanl.gov/abs/1110.6479}{{\tt arXiv:1110.6479}}].

\bibitem{Nollett:2011ho}
K.~M. {Nollett} and G.~P. {Holder}, {\it {An analysis of constraints on
  relativistic species from primordial nucleosynthesis and the cosmic microwave
  background}},  {\em ArXiv e-prints} (Dec., 2011)
  [\href{http://xxx.lanl.gov/abs/1112.2683}{{\tt arXiv:1112.2683}}].

\bibitem{Hou:2013da}
Z.~{Hou}, R.~{Keisler}, L.~{Knox}, M.~{Millea}, and C.~{Reichardt}, {\it {How
  massless neutrinos affect the cosmic microwave background damping tail}},
  {\em \prd} {\bf 87} (Apr., 2013) 083008.

\bibitem{Wagoner:1966pv}
R.~V. Wagoner, W.~A. Fowler, and F.~Hoyle, {\it {On the Synthesis of elements
  at very high temperatures}},  {\em Astrophys.J.} {\bf 148} (1967) 3--49.

\bibitem{Wagoner:1969sy}
R.~V. {Wagoner}, {\it {Synthesis of the Elements Within Objects Exploding from
  Very High Temperatures}},  {\em \apjs} {\bf 18} (June, 1969) 247.

\bibitem{SMK:1993bb}
M.~S. {Smith}, L.~H. {Kawano}, and R.~A. {Malaney}, {\it {Experimental,
  computational, and observational analysis of primordial nucleosynthesis}},
  {\em \apjs} {\bf 85} (Apr., 1993) 219--247.

\bibitem{Dolgov:1997ne}
A.~D. {Dolgov}, S.~H. {Hansen}, and D.~V. {Semikoz}, {\it {Non-equilibrium
  corrections to the spectra of massless neutrinos in the early universe}},
  {\em Nuclear Physics B} {\bf 503} (Feb., 1997) 426--444,
  [\href{http://xxx.lanl.gov/abs/hep-ph/9703315}{{\tt hep-ph/9703315}}].

\bibitem{Gnedin:1998ne}
N.~Y. {Gnedin} and O.~Y. {Gnedin}, {\it {Cosmological Neutrino Background
  Revisited}},  {\em \apj} {\bf 509} (Dec., 1998) 11--15,
  [\href{http://xxx.lanl.gov/abs/astro-ph/9712199}{{\tt astro-ph/9712199}}].

\bibitem{Mangano:3.046}
G.~{Mangano}, G.~{Miele}, S.~{Pastor}, and M.~{Peloso}, {\it {A precision
  calculation of the effective number of cosmological neutrinos}},  {\em
  Physics Letters B} {\bf 534} (May, 2002) 8--16,
  [\href{http://xxx.lanl.gov/abs/astro-ph/0111408}{{\tt astro-ph/0111408}}].

\bibitem{Fuller:2010nn}
R.~N. {Boyd}, C.~R. {Brune}, G.~M. {Fuller}, and C.~J. {Smith}, {\it {New
  nuclear physics for big bang nucleosynthesis}},  {\em \prd} {\bf 82} (Nov.,
  2010) 105005.

\bibitem{2010PhRvD..82l5017F}
G.~M. {Fuller} and C.~J. {Smith}, {\it {Nuclear weak interaction rates in
  primordial nucleosynthesis}},  {\em \prd} {\bf 82} (Dec., 2010) 125017,
  [\href{http://xxx.lanl.gov/abs/1009.0277}{{\tt arXiv:1009.0277}}].

\bibitem{Trotta:2004bc}
R.~{Trotta} and S.~H. {Hansen}, {\it {Constraining the helium abundance with
  CMB data}},  {\em \prd} {\bf 69} (Jan., 2004) 023509.

\bibitem{Ichikawa:2006bc}
K.~Ichikawa, T.~Sekiguchi, and T.~Takahashi, {\it Primordial helium abundance
  from cmb: A constraint from recent observations and a forecast},  {\em Phys.
  Rev. D} {\bf 78} (Aug, 2008) 043509.

\bibitem{AliHamoud:2011hr}
Y.~{Ali-Ha{\"i}moud} and C.~M. {Hirata}, {\it {HyRec: A fast and highly
  accurate primordial hydrogen and helium recombination code}},  {\em \prd}
  {\bf 83} (Feb., 2011) 043513.

\bibitem{Chluba:2011cr}
J.~{Chluba} and R.~M. {Thomas}, {\it {Towards a complete treatment of the
  cosmological recombination problem}},  {\em \mnras} {\bf 412} (Apr., 2011)
  748--764.

\bibitem{SSS:2000rc}
S.~{Seager}, D.~D. {Sasselov}, and D.~{Scott}, {\it {How Exactly Did the
  Universe Become Neutral?}},  {\em \apjs} {\bf 128} (June, 2000) 407--430.

\bibitem{SSS:1999nc}
S.~{Seager}, D.~D. {Sasselov}, and D.~{Scott}, {\it {A New Calculation of the
  Recombination Epoch}},  {\em \apjl} {\bf 523} (Sept., 1999) L1--L5.

\bibitem{1968ApJ...153....1P}
P.~J.~E. {Peebles}, {\it {Recombination of the Primeval Plasma}},  {\em \apj}
  {\bf 153} (July, 1968) 1.

\bibitem{1968ZhETF..55..278Z}
Y.~B. {Zeldovich}, V.~G. {Kurt}, and R.~A. {Syunyaev}, {\it {Recombination of
  Hydrogen in the Hot Model of the Universe}},  {\em Zhurnal Eksperimentalnoi i
  Teoreticheskoi Fiziki} {\bf 55} (July, 1968) 278--286.

\bibitem{1991A&A...251..680P}
D.~{Pequignot}, P.~{Petitjean}, and C.~{Boisson}, {\it {Total and effective
  radiative recombination coefficients}},  {\em \aap} {\bf 251} (Nov., 1991)
  680--688.

\bibitem{1998MNRAS.297.1073H}
D.~G. {Hummer} and P.~J. {Storey}, {\it {Recombination of helium-like ions - I.
  Photoionization cross-sections and total recombination and cooling
  coefficients for atomic helium}},  {\em \mnras} {\bf 297} (July, 1998)
  1073--1078.

\bibitem{Silk:1968dd}
J.~{Silk}, {\it {Cosmic Black-Body Radiation and Galaxy Formation}},  {\em
  \apj} {\bf 151} (Feb., 1968) 459.

\bibitem{Zaldarriaga:1995dd}
M.~{Zaldarriaga} and D.~D. {Harari}, {\it {Analytic approach to the
  polarization of the cosmic microwave background in flat and open universes}},
   {\em \prd} {\bf 52} (Sept., 1995) 3276--3287.

\bibitem{Hu:1997dd}
W.~{Hu} and M.~{White}, {\it {CMB anisotropies: Total angular momentum
  method}},  {\em \prd} {\bf 56} (July, 1997) 596--615.

\bibitem{Weinberg:2008co}
S.~Weinberg, {\em Cosmology}.
\newblock Oxford University Press, 2008.

\bibitem{KF:2008PRL}
G.~M. {Fuller} and C.~T. {Kishimoto}, {\it {Quantum Coherence of Relic
  Neutrinos}},  {\em Physical Review Letters} {\bf 102} (May, 2009) 201303,
  [\href{http://xxx.lanl.gov/abs/0811.4370}{{\tt arXiv:0811.4370}}].

\bibitem{GFP-PRL:2014mn}
E.~Grohs, G.~M. Fuller, C.~T. Kishimoto, and M.~W. Paris, {\it {Effects of
  $N_\textrm{eff}$ and neutrino rest mass on ionization equilibrium
  freeze-out}},  {\em ArXiv e-prints} (Dec., 2014)
  [\href{http://xxx.lanl.gov/abs/1412.6875}{{\tt arXiv:1412.6875}}].

\bibitem{2014SPIE.9153E..1PB}
B.~A. {Benson} et~al., {\it {SPT-3G: a next-generation cosmic microwave
  background polarization experiment on the South Pole telescope}},  in {\em
  Society of Photo-Optical Instrumentation Engineers (SPIE) Conference Series},
  vol.~9153 of {\em Society of Photo-Optical Instrumentation Engineers (SPIE)
  Conference Series}, p.~1, July, 2014.
\newblock \href{http://xxx.lanl.gov/abs/1407.2973}{{\tt arXiv:1407.2973}}.

\bibitem{2014ApJ...794..171T}
{The Polarbear Collaboration: P.~A.~R.~Ade} et~al., {\it {A Measurement of the
  Cosmic Microwave Background B-mode Polarization Power Spectrum at Sub-degree
  Scales with POLARBEAR}},  {\em \apj} {\bf 794} (Oct., 2014) 171,
  [\href{http://xxx.lanl.gov/abs/1403.2369}{{\tt arXiv:1403.2369}}].

\bibitem{TMT}
D.~Silva, P.~Hickson, C.~Steidel, and M.~Bolte, {\it {TMT Detailed Science
  Case: 2007}},  tech. rep., 2007.
\newblock http://www.tmt.org.

\bibitem{GMT}
P.~{McCarthy} and R.~A. {Bernstein}, {\it {Giant Magellan Telescope: Status and
  Opportunities for Scientific Synergy}},  in {\em Thirty Meter Telescope
  Science Forum}, p.~61, July, 2014.

\bibitem{EELT}
I.~{Hook}, {\em {The science case for the European Extremely Large Telescope :
  the next step in mankind's quest for the Universe}}.
\newblock 2005.

\bibitem{LSND:1997}
C.~{Athanassopoulos} et~al., {\it {The liquid scintillator neutrino detector
  and LAMPF neutrino source}},  {\em Nuclear Instruments and Methods in Physics
  Research A} {\bf 388} (Feb., 1997) 149--172,
  [\href{http://xxx.lanl.gov/abs/nucl-ex/9605002}{{\tt nucl-ex/9605002}}].

\bibitem{MiniBooNE:2010}
A.~A. {Aguilar-Arevalo} et~al., {\it {Measurement of the neutrino
  neutral-current elastic differential cross section}},  {\em \prd} {\bf 82}
  (Nov., 2010) 092005, [\href{http://xxx.lanl.gov/abs/1007.4730}{{\tt
  arXiv:1007.4730}}].

\bibitem{DoubleChooz:2012}
Y.~{Abe} et~al., {\it {Reactor $\bar{\nu_e}$ disappearance in the Double Chooz
  experiment}},  {\em \prd} {\bf 86} (Sept., 2012) 052008,
  [\href{http://xxx.lanl.gov/abs/1207.6632}{{\tt arXiv:1207.6632}}].

\bibitem{KF:2008PRD}
C.~T. {Kishimoto} and G.~M. {Fuller}, {\it {Lepton-number-driven sterile
  neutrino production in the early universe}},  {\em \prd} {\bf 78} (July,
  2008) 023524, [\href{http://xxx.lanl.gov/abs/0802.3377}{{\tt
  arXiv:0802.3377}}].

\bibitem{GFG:2015}
L.~A. Gilbert, G.~M. Fuller, and E.~Grohs in preparation, 2015.

\bibitem{2002NuPhB.632..363D}
A.~D. {Dolgov}, S.~H. {Hansen}, S.~{Pastor}, S.~T. {Petcov}, G.~G. {Raffelt},
  and D.~V. {Semikoz}, {\it {Cosmological bounds on neutrino degeneracy
  improved by flavor oscillations}},  {\em Nucl. Phys. B} {\bf 632} (June,
  2002) 363--382, [\href{http://xxx.lanl.gov/abs/hep-ph/0201287}{{\tt
  hep-ph/0201287}}].

\bibitem{DarkRadScherrer:2012}
J.~L. {Menestrina} and R.~J. {Scherrer}, {\it {Dark radiation from particle
  decays during big bang nucleosynthesis}},  {\em \prd} {\bf 85} (Feb., 2012)
  047301, [\href{http://xxx.lanl.gov/abs/1111.0605}{{\tt arXiv:1111.0605}}].

\bibitem{2008RPPh...71h6201O}
E.~W. {Otten} and C.~{Weinheimer}, {\it {Neutrino mass limit from tritium
  {$\beta$} decay}},  {\em Reports on Progress in Physics} {\bf 71} (Aug.,
  2008) 086201, [\href{http://xxx.lanl.gov/abs/0909.2104}{{\tt
  arXiv:0909.2104}}].

\bibitem{2008JPhCS.136b2031D}
G.~{Drexlin}, {\it {Direct neutrino mass measurements}},  {\em Journal of
  Physics Conference Series} {\bf 136} (Nov., 2008) 022031.

\bibitem{2005JHEP...11..028K}
A.~{Kusenko}, S.~{Pascoli}, and D.~{Semikoz}, {\it {Bounds on heavy sterile
  neutrinos revisited}},  {\em Journal of High Energy Physics} {\bf 11} (Nov.,
  2005) 28, [\href{http://xxx.lanl.gov/abs/hep-ph/0405198}{{\tt
  hep-ph/0405198}}].

\bibitem{1983PhLB..128..361B}
F.~{Bergsma} et~al., {\it {A search for decays of heavy neutrinos}},  {\em
  Physics Letters B} {\bf 128} (Sept., 1983) 361--366.

\bibitem{1986PhLB..166..479B}
G.~{Bernardi}, G.~{Carugno}, J.~{Chauveau}, F.~{Dicarlo}, M.~{Dris},
  J.~{Dumarchez}, M.~{Ferro-Luzzi}, J.-M. {Levy}, D.~{Lukas}, J.-M. {Perreau},
  Y.~{Pons}, A.-M. {Touchard}, and F.~{Vannucci}, {\it {Search for neutrino
  decay}},  {\em Physics Letters B} {\bf 166} (Jan., 1986) 479--483.

\bibitem{1988PhLB..203..332B}
G.~{Bernardi}, G.~{Carugno}, J.~{Chauveau}, F.~{Dicarlo}, M.~{Dris},
  J.~{Dumarchez}, M.~{Ferro-Luzzi}, J.-M. {Levy}, D.~{Lukas}, J.-M. {Perreau},
  Y.~{Pons}, A.-M. {Touchard}, and F.~{Vannucci}, {\it {Further limits on heavy
  neutrino couplings}},  {\em Physics Letters B} {\bf 203} (Mar., 1988)
  332--334.

\bibitem{1993PhLB..302..336B}
S.~{Baranov}, Y.~{Batusov}, S.~{Bunyatov}, O.~{Klimov}, V.~{Lyukov},
  Y.~{Nefedov}, B.~{Popov}, V.~{Valuev}, A.~{Borisov}, V.~{Goryachev},
  M.~{Kirsanov}, A.~{Kozhin}, V.~{Kravtsov}, A.~{Spiridonov}, V.~{Tumakov},
  A.~{Vovenko}, and D.~{Kiss}, {\it {Search for heavy neutrinos at the
  IHEP-JINR Neutrino Detector}},  {\em Physics Letters B} {\bf 302} (Mar.,
  1993) 336--340.

\bibitem{2001NuPhS..98...37N}
P.~{N{\'e}d{\'e}lec}, {\it {Tau-sterile neutrino mixing in NOMAD}},  {\em
  Nuclear Physics B Proceedings Supplements} {\bf 98} (Apr., 2001) 37--42.

\bibitem{2014arXiv1410.7684M}
S.~{Mertens}, K.~{Dolde}, M.~{Korzeczek}, F.~{Glueck}, S.~{Groh}, R.~D.
  {Martin}, A.~W.~P. {Poon}, and M.~{Steidl}, {\it {Wavelet Approach to Search
  for Sterile Neutrinos in Tritium \$$\backslash$beta\$-Decay Spectra}},  {\em
  ArXiv e-prints} (Oct., 2014) [\href{http://xxx.lanl.gov/abs/1410.7684}{{\tt
  arXiv:1410.7684}}].

\bibitem{2003PhRvL..91x1301K}
M.~{Kaplinghat}, L.~{Knox}, and Y.-S. {Song}, {\it {Determining Neutrino Mass
  from the Cosmic Microwave Background Alone}},  {\em Physical Review Letters}
  {\bf 91} (Dec., 2003) 241301,
  [\href{http://xxx.lanl.gov/abs/astro-ph/0303344}{{\tt astro-ph/0303344}}].

\bibitem{2004JCAP...04..008H}
S.~{Hannestad} and G.~{Raffelt}, {\it {Cosmological mass limits on neutrinos,
  axions, and other light particles}},  {\em \jcap} {\bf 4} (Apr., 2004) 8,
  [\href{http://xxx.lanl.gov/abs/hep-ph/0312154}{{\tt hep-ph/0312154}}].

\bibitem{2009PhRvD..80l3509D}
F.~{de Bernardis}, T.~D. {Kitching}, A.~{Heavens}, and A.~{Melchiorri}, {\it
  {Determining the neutrino mass hierarchy with cosmology}},  {\em \prd} {\bf
  80} (Dec., 2009) 123509, [\href{http://xxx.lanl.gov/abs/0907.1917}{{\tt
  arXiv:0907.1917}}].

\bibitem{2010JCAP...08..001H}
S.~{Hannestad}, A.~{Mirizzi}, G.~G. {Raffelt}, and Y.~Y.~Y. {Wong}, {\it
  {Neutrino and axion hot dark matter bounds after WMAP-7}},  {\em \jcap} {\bf
  8} (Aug., 2010) 1, [\href{http://xxx.lanl.gov/abs/1004.0695}{{\tt
  arXiv:1004.0695}}].

\bibitem{2010PhRvD..82h7302A}
M.~{Archidiacono}, A.~{Cooray}, A.~{Melchiorri}, and S.~{Pandolfi}, {\it {CMB
  neutrino mass bounds and reionization}},  {\em \prd} {\bf 82} (Oct., 2010)
  087302, [\href{http://xxx.lanl.gov/abs/1010.5757}{{\tt arXiv:1010.5757}}].

\bibitem{2011PhRvD..83k5023G}
E.~{Giusarma}, M.~{Corsi}, M.~{Archidiacono}, R.~{de Putter}, A.~{Melchiorri},
  O.~{Mena}, and S.~{Pandolfi}, {\it {Constraints on massive sterile neutrino
  species from current and future cosmological data}},  {\em \prd} {\bf 83}
  (June, 2011) 115023, [\href{http://xxx.lanl.gov/abs/1102.4774}{{\tt
  arXiv:1102.4774}}].

\bibitem{GFKPV-nuxport}
G.~M. Fuller, E.~Grohs, C.~T. Kishimoto, M.~W. Paris, and A.~Vlasenko in
  preparation, 2015.

\bibitem{1991ApJ...368....1S}
M.~J. {Savage}, R.~A. {Malaney}, and G.~M. {Fuller}, {\it {Neutrino
  oscillations and the leptonic charge of the universe}},  {\em \apj} {\bf 368}
  (Feb., 1991) 1--11.

\bibitem{2002PhRvD..66a3008A}
K.~N. {Abazajian}, J.~F. {Beacom}, and N.~F. {Bell}, {\it {Stringent
  constraints on cosmological neutrino-antineutrino asymmetries from
  synchronized flavor transformation}},  {\em \prd} {\bf 66} (July, 2002)
  013008, [\href{http://xxx.lanl.gov/abs/astro-ph/0203442}{{\tt
  astro-ph/0203442}}].

\bibitem{2002PhRvD..66b5015W}
Y.~Y. {Wong}, {\it {Analytical treatment of neutrino asymmetry equilibration
  from flavor oscillations in the early universe}},  {\em \prd} {\bf 66} (July,
  2002) 025015, [\href{http://xxx.lanl.gov/abs/hep-ph/0203180}{{\tt
  hep-ph/0203180}}].

\bibitem{2004NJPh....6..117K}
J.~P. {Kneller} and G.~{Steigman}, {\it {BBN for pedestrians}},  {\em New
  Journal of Physics} {\bf 6} (Sept., 2004) 117,
  [\href{http://xxx.lanl.gov/abs/astro-ph/0406320}{{\tt astro-ph/0406320}}].

\bibitem{2006PhRvD..74h5008S}
C.~J. {Smith}, G.~M. {Fuller}, C.~T. {Kishimoto}, and K.~N. {Abazajian}, {\it
  {Light element signatures of sterile neutrinos and cosmological lepton
  numbers}},  {\em \prd} {\bf 74} (Oct., 2006) 085008,
  [\href{http://xxx.lanl.gov/abs/astro-ph/0608377}{{\tt astro-ph/0608377}}].

\bibitem{AFG:2015}
J.~N. Ariche, G.~M. Fuller, and E.~Grohs in preparation, 2015.

\bibitem{Abazajian201566}
K.~Abazajian et~al., {\it Neutrino physics from the cosmic microwave background
  and large scale structure},  {\em Astroparticle Physics} {\bf 63} (2015),
  no.~0 66 -- 80. Dark Energy and \{CMB\}.

\end{thebibliography}\endgroup

\end{document}